\documentstyle[12pt,graphicx,subfigure]{article}
\def\Journal#1#2#3#4{{#1} {\bf #2}, #3 (#4)}

\def\PLB{{\em Phys. Lett.}  B}

\def\PRD{{\em Phys. Rev.} D}

\def\beq{\begin{equation}}
\def\eeq{\end{equation}}
\def\lsim{\ ^<\llap{$_\sim$}\ }
\def\gsim{\ ^>\llap{$_\sim$}\ }
\def\r2{\sqrt 2}
\def\beq{\begin{equation}}
\def\eeq{\end{equation}}
\def\beqn{\begin{eqnarray}}
\def\eeqn{\end{eqnarray}}

\def\sinW2{\sin^2\theta_W}

\def\mz2{M_{z}^2}
\def\c2b{\cos 2\beta}

\def\mz{M_z}

\def\Fq2{F_{2}(q^2)}

\def\sec2w{sec^2\theta_W}

\def\gmin2{(g-2)_\mu}

\catcode`\@=11 

\def\lsim{\mathrel{\mathpalette\@versim<}}
\def\gsim{\mathrel{\mathpalette\@versim>}}
\def\@versim#1#2{\vcenter{\offinterlineskip
    \ialign{$\m@th#1\hfil##\hfil$\crcr#2\crcr\sim\crcr } }}

\begin{document}

\begin{titlepage}

\begin{center}
{\large {\bf Effective Lagrangian for $\bar q\tilde q'_i\chi^+_j$,
$\bar q\tilde q'_i\chi^0_j$ interactions and Fermionic Decays of the
Squarks with CP Phases }}\\
\vskip 0.5 true cm
\vspace{2cm}
\renewcommand{\thefootnote}
{\fnsymbol{footnote}}
 Tarek Ibrahim$^{a,b}$ and  Pran Nath$^{b}$  
\vskip 0.5 true cm
\end{center}

\noindent
{a. Department of  Physics, Faculty of Science,
University of Alexandria,}\\
{ Alexandria, Egypt\footnote{: Permanent address of T.I.}}\\ 
{b. Department of Physics, Northeastern University,
Boston, MA 02115-5000, USA } \\
\vskip 1.0 true cm
\centerline{\bf Abstract}
\medskip
The one loop corrected effective Lagrangian for the quark-squark-chargino and 
quark-squark-neutralino interactions is computed. The effective Lagrangian takes into account the
loop corrections arising from the exchange of the gluinos, charginos, neutralinos, W, Z, 
the charged Higgs, and the neutral Higgs. 
We further analyze the squark decays into charginos and neutralinos and discuss the
effect of the loop corrections on them. The analysis takes into account 
CP phases in the soft parameters. It is found that the loop corrections to the stop decay 
widths into chargino and neutralinos  can be as much as thirty percent or even larger. 
Further, the stop decay widths show a strong dependence on the CP phases. 
These results are of relevance in the precision predictions of squark decays
in the context of specific models of soft breaking in supergravity and string based models.
\end{titlepage}

\section{Introduction}
The $\bar{q} \tilde{ q'_i} \chi^+_j$ and $\bar{q} \tilde{ q_i} \chi^0_j$ interactions are of great
interest since they enter in the decay of squarks.  We expect that such decays will be  observed 
at the collider experiments. Specifically one expects under the usual naturalness criteria 
that most of the sparticles should become visible at the Large Hadron Collider
(LHC) with the possibility that some of the sparticles 
may also become visible at  RUN II of the Tevatron.  
Measurements of sparticle masses and of their decay branching ratios will be
a primary focus of attention after the discovery of such particles, while a more precise
measurement will come eventually at the next linear collider (NLC). With the above in mind
it is of great importance to refine the theoretical computations of the decay branching 
ratios beyond the tree level predictions. In this paper we extend the previous analyzes
of squark decays into charginos and neutralinos\cite{bartl1,bartl2,aoki}  
 by taking into account the loop corrections with CP phases.
For this purpose it is found advantageous to  compute the one loop corrected  effective Lagrangian
for the $\bar{q} \tilde{ q'_i} \chi^+_j$ and $\bar{q} \tilde{ q_i} \chi^0_j$ couplings.
In the analysis we also include the effect of CP phases.  It is now well known that large CP  phases
can be made compatible\cite{na,incancel,olive,chang}  with the experimental constraints on 
the electric dipole moments of the
electron\cite{eedm}, of the neutron\cite{nedm}, and  of the $Hg^{199}$ atom\cite{atomic}. 
Further, if the phases are large they could affect a whole host
of low energy phenomena. These include the effect on the 
higgs masses, couplings and 
decays\cite{cphiggsmass,Christova:2002sw,Ibrahim:2003ca,Ibrahim:2003jm,Ibrahim:2003tq,ibrahim1,ibrahim2,ibrahim3}, 
dark matter\cite{cpdark,gomez}, and a variety of other phenomena\cite{Ibrahim:2002ry}.
The outline of the rest of  the paper is as follows:
In Sec.2 we compute the effective Lagrangian for the  $\bar{t} \tilde{ b_i} \chi^+_j$ and
 $\bar{b} \tilde{ t_i} \chi^C_j$ interactions.
In Sec.3 a similar analysis is done for the    $\bar{q} \tilde{ q_i} \chi^0_j$ interaction.
 In Sec.4 we give an analysis of the decay widths of the squarks into charginos and neutralinos
 using the effective Lagrangian.
In Sec.5 we give a numerical analysis of the size of the loop effects on the decay widths. We also study
in this section the effect of CP phases on the decay widths. Conclusions are given in Sec.6.

\section{Effective Lagrangian for $\bar q\tilde q'_i\chi^{\pm}_j$ Interaction}
In this section we study the effect of loop corrections on  $\bar t \tilde b_i \chi_j^+$
and on $\bar b \tilde t_i \chi_j^c$ interactions. We begin with the tree level Lagrangian density
\beqn
{\cal {L}}= g\bar t (R_{bij} P_R + L_{bij} P_L) \tilde \chi_j^+\tilde b_i\nonumber\\
+g\bar b (R_{tij}P_R +L_{tij}P_L) \tilde \chi_j^c \tilde t_i +H.c. 
\label{1}
\eeqn 
where
\beqn
R_{bij} = -(U_{j1} D_{b1i} -K_b U_{j2} D_{b2i})\nonumber\\
 L_{bij} = K_t V_{j2}^* D_{b1i}\nonumber\\
R_{tij}=-(V_{j1} D_{t1i} -K_t V_{j2} D_{t2i})\nonumber\\
 L_{tij}=K_b U_{j2}^* D_{t1i}
\label{2-5}
\eeqn
and where
\beqn
K_{t(b)}=\frac{m_{t(b)}}{\sqrt 2 m_W \sin\beta(\cos\beta)} 
\eeqn
and the matrices $U, V$ and $D_{b(t)}$ are the diagonalizing matrices of the chargino and
squark mass matrices so that 
\beqn
U^* M_{\chi^+}V^{-1}= diag (m_{\chi_1^+}, m_{\chi_2^+})\nonumber\\
D^{\dagger}_q M_{\tilde q}^2D_q= diag (m_{\tilde q_1}^2, m_{\tilde q_2}^2)
\eeqn
where $m_{\chi_i^+}$ (i=1,2) are the eigen values of the chargino mass
matrix  and $m_{\tilde q_I}^2$ (i=1,2) are the eigen values of the squark mass$^2$ 
matrix. 
The loop corrections produce shift in the couplings of Eq.~(\ref{2-5}) as follows
\beqn
{\cal L}_{eff}= g\bar t ((R_{bij} + \Delta R_{bij}) P_R + 
(L_{bij}+\Delta L_{bij})P_L) \tilde \chi_j^+\tilde b_i\nonumber\\
+g\bar b ( (R_{tij}+ \Delta R_{tij})  P_R + (L_{tij} + \Delta L_{tij})P_L) \tilde \chi_j^c \tilde t_i +H.c. 
\label{8}
\eeqn 
where $\Delta R_{bij}$,  $\Delta L_{bij}$, $\Delta R_{tij}$, and $\Delta L_{tij}$
are the corrections that arise from the diagrams in Figs.(1-4).
As is conventional we will use the zero external momentum approximation in the analysis 
of these corrections (see, e.g., Ref.\cite{carena2002}).  
\subsection{$\Delta R_{bij}$ and $\Delta L_{bij}$  analysis}

Contributions to $\Delta R_{bij}$ and $\Delta L_{bij}$ arise from the nine loop diagrams of 
Fig.(1). We discuss now in detail the contribution of each of these diagrams, figs. (1a-1i).  
We begin with the loop diagram of 
  Fig.(1a) which contributes the following to $\Delta R_{bij}$ and $\Delta L_{bij}$ 
\beqn
\Delta R_{bij}^{(1)}=\frac{2\alpha_s}{3\pi} \sum_{k=1}^2 K_b U_{j2}D_{t1k}^* D_{b1i} e^{i\xi_3} 
D_{t1k} m_b m_{\tilde g} f(m_b^2, m_{\tilde g}^2,m_{\tilde t_k}^2)
\label{9}
\eeqn
\beqn
\Delta L_{bij}^{(1)}=-\frac{2\alpha_s}{3\pi} \sum_{k=1}^2 (V_{j1}^*D_{t1k}^*- K_t V_{j2}^* D_{t2k}^*) 
D_{t2k} D_{b2i} e^{-i\xi_3} m_b  m_{\tilde g}
f(m_b^2, m_{\tilde g}^2,m_{\tilde t_k}^2)
\eeqn
where 
\beqn
f(x,y,z)= \frac{1}{(x-y)(x-z)(z-y)} 
(zxln\frac{z}{x} + xy ln\frac{x}{y} +yzln\frac{y}{z}) 
\label{11}
\eeqn
Next for the loop Fig.(1b) we find
\beqn
\Delta R_{bij}^{(2)}=
 \sum_{k=1}^2\sum_{l=1}^4
  2K_b U_{j2}D_{t1k}^*  (\beta_{bl}D_{b1i}+\alpha_{bl}^*D_{b2i}) 
  \nonumber\\  
  (\beta_{tl} D_{t1k} +\alpha_{tl}^* D_{t2k}) 
  \frac{m_bm_{\chi_l^0}}{16\pi^2} 
  f(m_b^2, m_{\chi_l^0}^2,m_{ \tilde t_k}^2)
\label{12}
\eeqn
\beqn
\Delta L_{bij}^{(2)}=-
 \sum_{k=1}^2\sum_{l=1}^4  
  2(V_{j1}^*D_{t1k}^*-K_t V_{j2}^*D_{t2k}^*) 
   (\alpha_{bl}D_{b1i}-\gamma_{bl}D_{b2i}) 
  \nonumber\\  
  (\alpha_{tl} D_{t1k} -\gamma_{tl} D_{t2k}) 
  \frac{m_bm_{\chi_l^0}}{16\pi^2} 
  f(m_b^2, m_{\chi_l^0}^2,m_{ \tilde t_k}^2)
\label{13}
\eeqn
\beqn
\alpha_{b(t)k} =\frac{g m_{b(t)}X_{3(4)k}}{2m_W\cos\beta(\sin\beta)}\nonumber\\
\beta_{b(t)k}=eQ_{b(t)}X_{1k}^{'*} +\frac{g}{\cos\theta_W} X_{2k}^{'*}
(T_{3b(t)}-Q_{b(t)}\sin^2\theta_W)\nonumber\\
\gamma_{b(t)k}=eQ_{b(t)} X_{1k}'-\frac{gQ_{b(t)}\sin^2\theta_W}{\cos\theta_W}
X_{2k}'
\eeqn
where $X'$'s are given by 
\beqn
X'_{1k}=X_{1k}\cos\theta_W +X_{2k}\sin\theta_W\nonumber\\
X'_{2k}=-X_{1k}\sin\theta_W +X_{2k}\cos\theta_W
\eeqn
and where X is the matrix that diagonalizes the neutralino mass matrix
so that 
\beqn
X^T M_{\chi^0}X= diag (m_{\chi_1^0},
m_{\chi_2^0},m_{\chi_3^0},m_{\chi_4^0})
\eeqn
Fig.(1c) contributes the following
\beqn
\Delta R_{bij}^{(3)} =\frac{1}{\sqrt{2}} 
\sum_{k=1}^2\sum_{l=1}^3 (U_{j1} D_{b1k} -k_b U_{j2}D_{b2k})\nonumber\\ 
(G_{ki}(Y_{l2}+iY_{l3} \cos\beta) +G_{ik}^* (Y_{l2}- iY_{l3} \cos\beta)
+H_{ki}(Y_{l1}+iY_{l3}\sin\beta)\nonumber\\
+H_{ik}^* (Y_{l1}-iY_{l3}\sin\beta))
(C^S_{tl}+iC^P_{tl})\frac{m_t}{16\pi^2} f(m_t^2, m_{\tilde b_k}^2,
m_{H_l}^2) 
\eeqn
where $Y$ is the diagonalizing matrix of the Higgs mass$^2$ matrix
\beqn
YM_{Higgs}^2Y^T =diag(m_{H_1}^2, m_{H_2}^2,m_{H_3}^2)
\eeqn
and 
\beqn
G_{ij}= \frac{gm_Z}{\sqrt 2 \cos\theta_W} 
 ((-\frac{1}{2} +\frac{1}{3}\sin^2\theta_W)D_{b1i}^*D_{b1j}
-\frac{1}{3}\sin^2\theta_W D_{b2i}^*D_{b2j})\sin\beta\nonumber\\ 
+\frac{gm_b}{\sqrt 2 m_W \cos\beta} \mu D_{b1i}^*D_{b2j}
\eeqn
\beqn
H_{ij}= -\frac{gm_Z}{\sqrt 2 \cos\theta_W} 
[(-\frac{1}{2} +\frac{1}{3}\sin^2\theta_W)D_{b1i}^*D_{b1j}
-\frac{1}{3}\sin^2\theta_W D_{b2i}^*D_{b2j}]\cos\beta\nonumber\\ 
-\frac{gm_b^2}{\sqrt 2 m_W \cos\beta} [ D_{b1i}^*D_{b1j}
+ D_{b2i}^*D_{b2j}] -\frac{gm_bm_0A_b}{\sqrt 2 m_W \cos\beta} 
D_{b2i}^*D_{b1j}
\eeqn
and
\beqn
C_{tl}^S= \tilde C_{tl}^S \cos\chi_t -\tilde C_{tl}^P \sin\chi_t\nonumber\\
 C_{tl}^P= \tilde C_{tl}^S \sin\chi_t +\tilde C_{tl}^P \cos\chi_t\nonumber\\
\sqrt 2 \tilde C_{tl}^S = Re (h_t +\delta h_t) Y_{l2} + [-Im (h_t + \delta h_t)
 \cos\beta\nonumber\\  + Im (\Delta h_t)\sin\beta ] Y_{l3}
+ Re (\Delta h_t) Y_{l1}\nonumber\\
\sqrt 2 \tilde C_{tl}^P = - Im (h_t +\delta h_t) Y_{l2} + [-Re (h_t + \delta h_t)
\cos\beta\nonumber\\ + Re (\Delta h_t) \sin\beta ]Y_{l3}
- Im (\Delta h_t) Y_{l1}
\eeqn
with
\beqn
\tan\chi_t =\frac{ Im (\frac{\delta h_t}{h_t}
+ \frac{\Delta h_t}{h_t}\cot\beta)}
{1+ Re (\frac{\delta h_t}{h_t}
+ \frac{\Delta h_t}{h_t}\cot\beta)}
\label{tanchib}
\eeqn
and
\beq
h_t=\frac{g m_t}{\sqrt{2} m_W \sin\beta}
\eeq
The corrections $\Delta h_t$ and $\delta h_t$ are defined in Appendix A.
\beqn
\Delta L_{bij}^{(3)} =-\frac{1}{\sqrt{2}} 
\sum_{k=1}^2 \sum_{l=1}^3 K_t V_{j2}^* D_{b1k}
 [G_{ki}(Y_{l2} +iY_{l3}\cos\beta) +G_{ik}^* (Y_{l2}-iY_{l3}
\cos\beta)\nonumber\\
+ H_{ki} (Y_{l1}+ iY_{l3} \sin\beta) +
H_{ik}^* (Y_{l1}-iY_{l3}\sin\beta)] (C^S_{tl}-iC^P_{tl})
\frac{m_t}{16\pi^2} f(m_t^2, m_{\tilde b_k}^2, m_{H_l}^2)
\label{23}
\eeqn
Fig.(1d) gives the following contributions 
\beqn
\Delta R_{bij}^{(4)} =\frac{\sqrt{2}}{g} 
\sum_{l=1}^4 (\beta_{bl}D_{b1i}+\alpha_{bl}^* D_{b2i})(B^{S*}_{bt}
-B^{P*}_{bt})
\xi_{lj}'\sin\beta \frac{m_bm_{\chi_l^0}} {16\pi^2} 
f(m_b^2, m_{\chi_l^0}^2, m_{H^-}^2)
\eeqn
where
\beqn
B_{bt}^S =-\frac{1}{2} (h_b + \overline{\delta h_b}) e^{-i\theta_{bt}}
\sin\beta + \frac{1}{2}  \overline{\Delta h_b} e^{-i\theta_{bt}}
\cos\beta\nonumber\\
-\frac{1}{2} (h_t + \overline{\delta h_t^*}) e^{i\theta_{bt}}
\cos\beta + \frac{1}{2}  \overline{\Delta h_t^*} e^{i\theta_{bt}}
\sin\beta\nonumber\\
B_{bt}^P =-\frac{1}{2} (h_t + \overline{\delta h_t^*}) e^{i\theta_{bt}}
\cos\beta + \frac{1}{2}  \overline{\Delta h_t^*} e^{i\theta_{bt}}
\sin\beta\nonumber\\
+\frac{1}{2} (h_b + \overline{\delta h_b}) e^{-i\theta_{bt}}
\sin\beta - \frac{1}{2}  \overline{\Delta h_b} e^{-i\theta_{bt}}
\cos\beta
\eeqn
where $\theta_{bt}=(\chi_b+\chi_t)/2$ and where
$\chi_b$ is  defined by the following
\beqn
\tan\chi_b =\frac{ Im (\frac{\delta h_b}{h_b}
+ \frac{\Delta h_b}{h_b}\tan\beta)}
{1+ Re (\frac{\delta h_b}{h_b}
+ \frac{\Delta h_b}{h_b}\tan\beta)}
\label{tanchib}
\eeqn
and 
\beq
h_b=\frac{g m_b}{\sqrt{2} m_W \cos\beta}
\eeq
where the corrections $\Delta h_f$, $\delta h_f$, $\bar{\Delta h_f}$ and
$\bar{\delta h_f}$ are defined in Appendix A.
\beqn
\Delta L_{bij}^{(4)} =\frac{\sqrt{2}}{g} 
\sum_{l=1}^4 (\alpha_{bl}D_{b1i}-\gamma_{bl} D_{b2i})(B^{S*}_{bt}
+B^{P*}_{bt})
\xi_{lj}\cos\beta \frac{m_bm_{\chi_l^0}} {16\pi^2} 
f(m_b^2, m_{\chi_l^0}^2, m_{H^-}^2)
\eeqn
where 
\beqn
\xi_{ji}'= -gX_{3j}^* U_{i1} +\frac{g}{\sqrt 2} X_{2j}^* U_{i2} 
+\frac{1}{\sqrt 2} g\tan\theta_W X_{1j}^* U_{i2}\nonumber\\
\xi_{ji}= -gX_{4j} V_{i1}^* -\frac{g}{\sqrt 2} X_{2j} V_{i2}^*
-\frac{1}{\sqrt 2} g\tan\theta_W X_{1j} V_{i2}^*\nonumber\\
\label{26}
\eeqn
Next we discuss the contributions from Fig.(1e). Here  on using
the properties of the projection operators, i.e.,  
$\gamma^{\mu}P_R=P_L \gamma^{\mu}$, $P_LP_R=0$,  and the property of 
Dirac $\gamma^{\mu}$ that $g_{\mu\nu}\gamma^{\mu}\gamma^{\nu}=4$,  we get 
\beqn
\Delta R_{bij}^{(5)}=0
\eeqn
and 
\beqn
\Delta L_{bij}^{(5)} =-4g 
\sum_{l=1}^4  R_{lj}'(\alpha_{bl}D_{b1i}-\gamma_{bl} D_{b2i}) 
 \frac{m_bm_{\chi_l^0}} {16\pi^2} 
f(m_b^2, m_{\chi_l^0}^2, m_{W^-}^2)
\eeqn
where
\beqn
R_{ij}'=\frac{1}{\sqrt 2} X_{3i}U_{j3} +X_{2i}U_{j1}
\eeqn
Contributions from Fig.(1f) are as follows
\beqn
\Delta R_{bij}^{(6)} =-g 
\sum_{l=1}^2\sum_{k=1}^3 [Q_{jl}(Y_{k1}+ iY_{k3}\sin\beta) 
+S_{jl}(Y_{k2}+i Y_{k3}\cos\beta)]\nonumber\\ 
(C^S_{tk}+i C^P_{tk}) 
[U_{l1}D_{b1i}-K_b U_{l2}D_{b2i}] 
\frac{m_tm_{\chi_l^-}}{16\pi^2} 
f(m_t^2, m_{\chi_l^-}^2, m_{H_k}^2)
\eeqn
and
\beqn
\Delta L_{bij}^{(6)} =g 
\sum_{l=1}^2\sum_{k=1}^3 [Q_{lj}^*(Y_{k1}- iY_{k3}\sin\beta) 
+S_{lj}^*(Y_{k2}-i Y_{k3}\cos\beta)]\nonumber\\ 
(C^S_{tk}-i C^P_{tk}) (K_tV_{l2}^* D_{b1i})  
\frac{m_tm_{\chi_l^-}}{16\pi^2} 
f(m_t^2, m_{\chi_l^-}^2, m_{H_k}^2)
\eeqn
where 
\beqn
Q_{ij}=\frac{1}{\sqrt 2} U_{i2}V_{j1}\nonumber\\
S_{ij}=\frac{1}{\sqrt 2} U_{i1}V_{j2}
\eeqn
Fig.(1g) contributes as follows
\beqn
\Delta R_{bij}^{(7)} =\frac{4g^2}{\cos^2\theta_W} 
\sum_{l=1}^2 L_{lj}^{''}(\frac{2}{3}\sin^2\theta_W) 
( U_{l1}D_{b1i}-K_bU_{l2}D_{b2i})\nonumber\\ 
\frac{m_tm_{\chi_l^-}}{16\pi^2} 
f(m_t^2, m_{\chi_l^-}^2, m_Z^2)
\eeqn
\beqn
\Delta L_{bij}^{(7)} =\frac{4g^2}{\cos^2\theta_W} 
\sum_{l=1}^2 R_{lj}^{''}(\frac{1}{2}- \frac{2}{3}\sin^2\theta_W) 
K_t V_{l2}^* D_{b1i}
\frac{m_tm_{\chi_l^-}}{16\pi^2} 
f(m_t^2, m_{\chi_l^-}^2, m_Z^2)
\eeqn
where
\beqn
L_{ij}^{''}=-V_{i1}V_{j1}^* -\frac{1}{2} V_{i2}V_{j2}^* 
+\delta_{ij} \sin^2\theta_W\nonumber\\
R_{ij}^{''}=-U_{i1}^*U_{j1} -\frac{1}{2} U^*_{i2}U_{j2} 
+\delta_{ij} \sin^2\theta_W
\eeqn
The contribution of Fig.(1h) is as follows
\beqn
\Delta R_{bij}^{(8)} =-\frac{\sqrt 2}{g} 
\sum_{l=1}^2\sum_{k=1}^4 \xi_{kj}' \sin\beta 
(\beta_{tk} D_{t1l} +\alpha_{tk}^* D_{t2l})\nonumber\\
(\eta_{il}\cos\beta +\eta_{il}^{'*}\sin\beta)   
\frac{m_{\chi_k^0}}{16\pi^2} 
f(m_{\chi_k^0}^2, m_{H^-}^2, m_{\tilde t_l}^2)
\eeqn
and
\beqn
\Delta L_{bij}^{(8)} =-\frac{\sqrt 2}{g} 
\sum_{l=1}^2\sum_{k=1}^4 \xi_{kj} \cos\beta 
(\alpha_{tk} D_{t1l} -\gamma_{tk} D_{t2l})\nonumber\\
(\eta_{il}\cos\beta +\eta_{il}^{'*}\sin\beta)   
\frac{m_{\chi_k^0}}{16\pi^2} 
f(m_{\chi_k^0}^2, m_{H^-}^2, m_{\tilde t_l}^2)
\eeqn
where
\beqn
\eta_{ij}= \frac{gm_t}{\sqrt 2m_W\sin\beta} m_0A_t D_{b1i} D_{t2j}^* 
+ \frac{gm_b}{\sqrt 2m_W\cos\beta} \mu  D_{b2i} D_{t1j}^*\nonumber\\
+ \frac{gm_bm_t}{\sqrt 2m_W\sin\beta}  D_{b2i} D_{t2j}^* 
+ \frac{gm_t^2}{\sqrt 2m_W\sin\beta}  D_{b1i} D_{t1j}^*
-\frac{g}{\sqrt 2} m_W \sin\beta D_{b1i}D_{t1j}^* 
\eeqn
and 
\beqn
\eta_{ji}'= \frac{gm_b}{\sqrt 2m_W\cos\beta} m_0A_b D_{b2j}^* D_{t1i} 
+ \frac{gm_t}{\sqrt 2m_W\sin\beta} \mu  D_{b1j}^* D_{t2i}\nonumber\\
+ \frac{gm_bm_t}{\sqrt 2m_W\cos\beta}  D_{b2j}^* D_{t2i} 
+ \frac{gm_b^2}{\sqrt 2m_W\cos\beta}  D_{b1j}^* D_{t1i}
-\frac{g}{\sqrt 2} m_W \cos\beta D_{b1j}^*D_{t1i} 
\eeqn
Finally the contribution from Fig.(1i) is as follows
\beqn
\Delta R_{bij}^{(9)} 
=\frac{g}{\sqrt 2} 
\sum_{l=1}^3 \sum_{s=1}^2\sum_{k=1}^2    
\{ Q_{js} (Y_{l1} + iY_{l3}\sin\beta) +S_{js} (Y_{l2} + iY_{l3} \cos\beta)\}
\nonumber\\
 (U_{s1}D_{b1k}-K_b U_{s2}D_{b2k})  
 [ G_{ki}(Y_{l2} + iY_{l3} \cos\beta)  + G_{ik}^* (Y_{l2}-iY_{ls}\cos\beta) 
 +H_{ki} (Y_{l1} + iY_{l3}\sin\beta)\nonumber\\ 
 + H_{ik}^* (Y_{l1}-iY_{l3}\sin\beta)] 
 \frac{m_{\chi_s^-}}{16\pi^2} 
f(m_{\tilde b_k}^2, m_{H_l}^2, m_{\chi_s^-}^2)
\eeqn
\beqn
\Delta L_{bij}^{(9)} 
=-\frac{g}{\sqrt 2} 
\sum_{l=1}^3 \sum_{s=1}^2\sum_{k=1}^2    
\{ Q_{sj}^* (Y_{l1} - iY_{l3}\sin\beta) +S_{sj}^* (Y_{l2} - iY_{l3} \cos\beta)\}
\nonumber\\
 (K_t V_{s2}^*D_{b1k})  
 [ G_{ki}(Y_{l2} + iY_{l3} \cos\beta)  + G_{ik}^* (Y_{l2}-iY_{l3}\cos\beta) 
 +H_{ki} (Y_{l1} + iY_{l3}\sin\beta)\nonumber\\ 
 + H_{ik}^* (Y_{l1}-iY_{l3}\sin\beta)] 
 \frac{m_{\chi_s^-}}{16\pi^2} 
f(m_{\tilde b_k}^2, m_{H_l}^2, m_{\chi_s^-}^2)
\label{42}
\eeqn
Summing the contribution from the nine loop diagrams of Fig.(1a-1i) we find that 
$\Delta R_{bj}$ and $\Delta L_{bij}$ that appear in Eq.~(\ref{8}) are then given
by 
\beqn
\Delta R_{bij} = \sum_{n=1}^9 \Delta R_{bij}^{(n)}
\label{43}
\eeqn
\beqn
\Delta L_{bij} = \sum_{n=1}^9 \Delta L_{bij}^{(n)}
\label{44}
\eeqn
We note  that the loop diagram of Fig.(1c) with the interchange 
$Z\leftrightarrow  H_l^0$
vanishes in the zero external momentum approximation because
the vertex is proportional to the external momentum, 
Similarly, the loop diagram of Fig.(1h) with the interchange $W^-\leftrightarrow   H^-$
vanishes and the loop diagram of Fig.(1i) with the interchange $Z\leftrightarrow  H_l^0$
vanishes in the zero external momentum approximation.
\subsection{Analysis of corrections $\Delta R_{tij}$ and  $\Delta L_{tij}$ }
The corrections $\Delta R_{tij}$ and  $\Delta L_{tij}$ arise from the nine diagrams 
of Fig.(2), i.e., the loops (a)-(i) of Fig.(2). We label the contribution from the
nine diagrams by superscripts 1-9. Thus, for example, the contributions of Fig.(2a)
are $\Delta R_{tij}^{(1)}$ and $\Delta L_{bij}^{(1)}$ etc.We now list
the contributions of the nine loops of Fig.(2). We have
\beqn 
\Delta R_{tij}^{(1)}=\frac{2\alpha_s}{3\pi} \sum_{k=1}^2 K_t V_{j2}D_{b1k}^* D_{t_1i} e^{i\xi_3} 
D_{b1k} m_t m_{\tilde g} f(m_t^2, m_{\tilde g}^2,m_{\tilde b_k}^2)
\label{45}
\eeqn
\beqn
\Delta L_{tij}^{(1)}=-\frac{2\alpha_s}{3\pi} \sum_{k=1}^2 (U_{j1}^*D_{b1k}^*- K_b U_{j2}^* D_{b2k}^*)  
D_{t2i} D_{b2k} e^{-i\xi_3} m_t m_{\tilde g}
f(m_t^2, m_{\tilde g}^2,m_{\tilde b_k}^2)
\label{46}
\eeqn
\beqn
\Delta R_{tij}^{(2)}=
 \sum_{k=1}^2\sum_{l=1}^4
  2K_t V_{j2}D_{b1k}^* (\beta_{tl}D_{t1i}+\alpha_{tl}^*D_{t2i}) 
  \nonumber\\   
  (\beta_{bl} D_{b1k} +\alpha_{bl}^* D_{b2k}) 
  \frac{m_bm_{\chi_l^0}}{16\pi^2}  
  f(m_b^2, m_{\chi_l^0}^2,m_{ \tilde t_k}^2)
\label{47}
\eeqn
\beqn
\Delta L_{tij}^{(2)}=-
 \sum_{k=1}^2\sum_{l=1}^4  
  2(U_{j1}^*D_{b1k}^*-K_b U_{j2}^*D_{b2k}^*) 
   (\alpha_{tl}D_{t1i}-\gamma_{tl}D_{t2i}) 
  \nonumber\\  
  (\alpha_{bk} D_{b1k} -\gamma_{bl} D_{b2k}) 
  \frac{m_bm_{\chi_l^0}}{16\pi^2} 
  f(m_b^2, m_{\chi_l^0}^2,m_{ \tilde t_k}^2)
\label{48}
\eeqn
\beqn
\Delta R_{tij}^{(3)} =\frac{1}{\sqrt{2}} 
\sum_{k=1}^2\sum_{l=1}^3 (V_{j1} D_{t1k} -K_t V_{j2}D_{t2k})\nonumber\\ 
(E_{ki}(Y_{l2}+iY_{l3} \cos\beta) +E_{ik}^* (Y_{l2}- iY_{l3} \cos\beta)
+F_{ki}(Y_{l1}+iY_{l3}\sin\beta)\nonumber\\
+F_{ik}^* (Y_{l1}-iY_{l3}\sin\beta))
(C^S_{bl}+i C^P_{bl})
\frac{m_b}{16\pi^2} f(m_t^2, m_{\tilde t_k}^2,
m_{H_l}^2) 
\eeqn
where 
\beqn
E_{ij}= \frac{gm_Z}{\sqrt 2 \cos\theta_W} 
 ((\frac{1}{2} -\frac{2}{3}\sin^2\theta_W)D_{t1i}^*D_{t1j}
+\frac{2}{3}\sin^2\theta_W D_{t2i}^*D_{t2j})\sin\beta\nonumber\\ 
-\frac{gm_t^2}{\sqrt 2 m_W \sin\beta} [D_{t1i}^* D_{t1j} +D_{t2i}^* D_{t2j}]
-\frac{gm_t m_0A_t}{\sqrt 2 m_W \sin\beta}D_{t2i}^*D_{t1j}
\eeqn
\beqn
F_{ij}= -\frac{gm_Z}{\sqrt 2 \cos\theta_W} 
[(\frac{1}{2} -\frac{2}{3}\sin^2\theta_W)D_{t1i}^*D_{t1j}
+\frac{2}{3}\sin^2\theta_W D_{t2i}^*D_{t2j}]\cos\beta\nonumber\\ 
 +\frac{gm_t \mu}{\sqrt 2 m_W \sin\beta} 
D_{t1i}^*D_{t2j}
\eeqn
and
\beqn
C_{bl}^S= \tilde C_{bl}^S \cos\chi_b -\tilde C_{bl}^P \sin\chi_b\nonumber\\
 C_{bl}^P= \tilde C_{bl}^S \sin\chi_b +\tilde C_{bl}^P \cos\chi_b\nonumber\\
\sqrt 2 \tilde C_{bl}^S = Re (h_b +\delta h_b) Y_{l1} + [-Im (h_b + \delta h_b)
 \sin\beta\nonumber\\  + Im (\Delta h_b)\cos\beta ] Y_{l3}
+ Re (\Delta h_b) Y_{l2}\nonumber\\
\sqrt 2 \tilde C_{bl}^P = - Im (h_b +\delta h_b) Y_{l1} + [-Re (h_b + \delta h_b)
\sin\beta\nonumber\\ + Re (\Delta h_b) \cos\beta ]Y_{l3}
- Im (\Delta h_t) Y_{l2}
\eeqn
\beqn
\Delta L_{tij}^{(3)} =-\frac{1}{\sqrt{2}} 
\sum_{k=1}^2 \sum_{l=1}^3 K_b U_{j2}^* D_{t1k}
 [E_{ki}(Y_{l2} +iY_{l3}\cos\beta) +E_{ik}^* (Y_{l2}-iY_{l3}
\cos\beta)\nonumber\\
+ F_{ki} (Y_{l1}+ iY_{l3} \sin\beta)+ 
F_{ik}^* (Y_{l1}-iY_{l3}\sin\beta)] (C^S_{bl}-i C^P_{bl})
\frac{m_b}{16\pi^2} f(m_b^2, m_{\tilde t_k}^2, m_{H_l}^2)
\label{52}
\eeqn
\beqn
\Delta R_{tij}^{(4)} =\frac{\sqrt{2}}{g} 
\sum_{l=1}^4 (\beta_{tl}D_{t1i}+\alpha_{tl}^* D_{t2i})(B^S_{bt}+B^P_{bt})
\xi_{lj}^*\cos\beta \frac{m_tm_{\chi_l^0}} {16\pi^2} 
f(m_t^2, m_{\chi_l^0}^2, m_{H^-}^2)
\eeqn
\beqn
\Delta L_{tij}^{(4)} =\frac{\sqrt{2}}{g}
\sum_{l=1}^4 (\alpha_{tl}D_{t1i}-\gamma_{tl} D_{t2i})(B^S_{bt}-B^P_{bt})
\xi_{lj}^{'*}\sin\beta \frac{m_tm_{\chi_l^0}} {16\pi^2} 
f(m_t^2, m_{\chi_l^0}^2, m_{H^-}^2)
\eeqn
\beqn
\Delta R_{tij}^{(5)}=0
\eeqn
and 
\beqn
\Delta L_{tij}^{(5)} =-4g 
\sum_{l=1}^4  R_{lj}'^*(\alpha_{tl}D_{t1i}-\gamma_{tl} D_{t2i}) 
 \frac{m_tm_{\chi_l^0}} {16\pi^2} 
f(m_t^2, m_{\chi_l^0}^2, m_{W^-}^2)
\label{56}
\eeqn
\beqn
\Delta R_{tij}^{(6)} =-g
\sum_{l=1}^2\sum_{k=1}^3 [Q_{lj}(Y_{k1}+ iY_{k3}\sin\beta) 
+S_{lj}(Y_{k2}+i Y_{k3}\cos\beta)]\nonumber\\ 
(C^S_{bk}+iC^P_{bk}) 
[V_{l1}D_{t1i}-K_t V_{l2}D_{t2i}] 
\frac{m_bm_{\chi_l^-}}{16\pi^2} 
f(m_b^2, m_{\chi_l^-}^2, m_{H_k}^2)
\label{57}
\eeqn
and
\beqn
\Delta L_{tij}^{(6)} =g 
\sum_{l=1}^2\sum_{k=1}^3 [Q_{jl}^*(Y_{k1}- iY_{k3}\sin\beta) 
+S_{jl}^*(Y_{k2}-i Y_{k3}\cos\beta)]\nonumber\\ 
(C^S_{bk}-iC^P_{bk})
 (K_bU_{l2}^* D_{t1i})  
\frac{m_bm_{\chi_l^-}}{16\pi^2} 
f(m_b^2, m_{\chi_l^-}^2, m_{H_k}^2)
\label{58}
\eeqn
\beqn
\Delta R_{tij}^{(7)} =-\frac{4g^2}{\cos^2\theta_W} 
\sum_{l=1}^2 L_{jl}^{''}(\frac{1}{3}\sin^2\theta_W) 
( V_{l1}D_{t1i}-K_tV_{l2}D_{t2i})\nonumber\\ 
\frac{m_bm_{\chi_l^-}}{16\pi^2} 
f(m_b^2, m_{\chi_l^-}^2, m_Z^2)
\label{59}
\eeqn
\beqn
\Delta L_{tij}^{(7)} =-\frac{4g^2}{\cos^2\theta_W} 
\sum_{l=1}^2 R_{jl}^{''}(\frac{1}{2}- \frac{1}{3}\sin^2\theta_W) 
K_b U_{l2}^* D_{t1i}
\frac{m_tm_{\chi_l^-}}{16\pi^2} 
f(m_b^2, m_{\chi_l^-}^2, m_Z^2)
\label{60}
\eeqn
where
\beqn
\Delta R_{tij}^{(8)} =-\frac{\sqrt 2}{g} 
\sum_{l=1}^2\sum_{k=1}^4 \xi_{jk}^* \cos\beta 
(\beta_{bk} D_{b1l} +\alpha_{bk}^* D_{b2l})\nonumber\\
(\eta_{li}'\sin\beta +\eta_{li}^{'*}\cos\beta)   
\frac{m_{\chi_k^0}}{16\pi^2} 
f(m_{\chi_k^0}^2, m_{H^-}^2, m_{\tilde b_l}^2)
\label{61}
\eeqn
and
\beqn
\Delta L_{tij}^{(8)} =-\frac{\sqrt 2}{g} 
\sum_{l=1}^2\sum_{k=1}^4 \xi_{jk}^{'*} \sin\beta 
(\alpha_{bk} D_{b1l} -\gamma_{bk} D_{b2l})\nonumber\\
(\eta_{li}'\sin\beta +\eta_{li}^*\cos\beta)   
\frac{m_{\chi_k^0}}{16\pi^2} 
f(m_{\chi_k^0}^2, m_{H^-}^2, m_{\tilde b_l}^2)
\label{62}
\eeqn
\beqn
\Delta R_{tij}^{(9)} 
=\frac{g}{\sqrt 2} 
\sum_{l=1}^3 \sum_{s=1}^2\sum_{k=1}^2    
\{ Q_{sj} (Y_{l1} + iY_{l3}\sin\beta) +S_{sj} (Y_{l2} + iY_{l3} \cos\beta)\}
\nonumber\\
 (V_{s1}D_{t1k}-K_t V_{s2}D_{t2k})  
 [ E_{ki}(Y_{l2} + iY_{l3} \cos\beta)  + E_{ik}^* (Y_{l2}-iY_{l3}\cos\beta) 
 +F_{ki} (Y_{l1} + iY_{l3}\sin\beta)\nonumber\\ 
 + F_{ik}^* (Y_{l1}-iY_{l3}\sin\beta)] 
 \frac{m_{\chi_s^-}}{16\pi^2} 
f(m_{\tilde t_k}^2, m_{H_l}^2, m_{\chi_s^-}^2)
\label{63}
\eeqn
\beqn
\Delta L_{bij}^{(9)} 
=-\frac{g}{\sqrt 2} 
\sum_{l=1}^3 \sum_{s=1}^2\sum_{k=1}^2    
\{ Q_{js}^* (Y_{l1} - iY_{l3}\sin\beta) +S_{js}^* (Y_{l2} - iY_{l3} \cos\beta)\}
\nonumber\\
 (K_b U_{s2}^*D_{t1k})  
 [ E_{ki}(Y_{l2} + iY_{l3} \cos\beta)  + E_{ik}^* (Y_{l2}-iY_{l3}\cos\beta) 
 +F_{ki} (Y_{l1} + iY_{l3}\sin\beta)\nonumber\\ 
 + F_{ik}^* (Y_{l1}-iY_{l3}\sin\beta)] 
 \frac{m_{\chi_s^-}}{16\pi^2} 
f(m_{\tilde t_k}^2, m_{H_l}^2, m_{\chi_s^-}^2)
\label{64}
\eeqn
\beqn
\Delta R_{tij} = \sum_{n=1}^9 \Delta R_{tij}^{(n)}
\label{65}
\eeqn
\beqn
\Delta L_{tij} = \sum_{n=1}^9 \Delta L_{tij}^{(n)}
\label{65}
\eeqn
The loops corresponding to Fig.(2c) and Fig.(2i) where $Z\leftrightarrow H_i^0$ 
vanishes for the same  reason as discussed earlier in Sec.(2.1). Similarly,
the loop corresponding to Fig.(2h) with $W^+\leftrightarrow H^+$ 
vanishes for  the same reason.
\section{The effective Lagrangian for $\bar q \tilde q_i \chi_j^0$ 
interaction} 
We turn now to an analysis of the loop corrections  to the squark-quark-neutralino 
interaction. We begin with the tree level $\bar q \tilde q_i \chi_j^0$  interaction 
which is given by   
\beqn
{\cal {L}}= g\bar b [K_{bij} P_R + M_{bij} P_L]\chi_j^0 \tilde b_i\nonumber\\ 
+g \bar t [K_{tij} P_R + M_{tij} P_L]\chi_j^0 \tilde t_i + H.c. 
\eeqn
where
\beqn
K_{bij}= -\sqrt 2 [\beta_{bj}D_{b1i} + \alpha_{bj}^* D_{b2i}]  \nonumber\\
K_{tij}= -\sqrt 2 [\beta_{tj}D_{t1i} + \alpha_{tj}^* D_{t2i}]  \nonumber\\
M_{bij}= -\sqrt 2 [\alpha_{bj}D_{b1i} - \gamma_{bj} D_{b2i}]  \nonumber\\
M_{tij}= -\sqrt 2 [\alpha_{tj}D_{t1i} - \gamma_{tj} D_{t2i}]  
\label{68-71}
\eeqn
The loop corrections  produce a shift in the couplings of  Eq.(\ref{68-71}) as follows
\beqn
{\cal {L}}_{eff}= g\bar b [(K_{bij} + \Delta K_{bij})   P_R +
 (M_{bij} + \Delta M_{bij}) P_L]\chi_j^0 \tilde b_i\nonumber\\ 
+g \bar t [  (K_{tij}+\Delta  K_{tij})P_R + 
(M_{tij} + \Delta M_{tij}) P_L]\chi_j^0 \tilde t_i + H.c. 
\label{72}
\eeqn
Thus in this part of the analysis we will calculate the quantities 
$\Delta K_{bij}$, $\Delta K_{tij}$, $\Delta M_{bij}$, and $\Delta M_{tij}$
from the one loop corrections arising from Figs.(3) and (4) using as in the previous anslysis
the zero external momentum approximation.
 \subsection{ Analysis of loop corrections to $\bar b\tilde b_i \chi_j^0$ interaction}
Loop corrections to $\bar b\tilde b_i \chi_j^0$ interaction, i.e.,  
$\Delta K_{bij}$ and $\Delta M_{bij}$,  arise from the nine diagrams of Fig.(3). 
We give now the individual contribution of these  nine loops. 
The contribution from Fig.(3a) is 
\beqn
\Delta K_{bij}^{(1)}=-\frac{2\sqrt 2 \alpha_s}{3\pi g}
\sum_{k=1}^2 e^{i\xi_3} D_{b1i}D_{b1k} 
(\alpha_{bj}^*D_{b1k}^* - \gamma_{bj}^* D_{b2k}^*)
m_b m_{\tilde g} f(m_b^2, m_{\tilde g}^2, m_{b_k}^2)
\label{73}
\eeqn
\beqn
\Delta M_{bij}^{(1)}=-\frac{2\sqrt 2 \alpha_s}{3\pi g}
\sum_{k=1}^2 e^{-i\xi_3} D_{b2i}D_{b2k} 
(\beta_{bj}^*D_{b1k}^* + \alpha_{bj} D_{b2k}^*)
m_b m_{\tilde g} f(m_b^2, m_{\tilde g}^2, m_{b_k}^2)
\label{74}
\eeqn
Fig.(3b) contibutes as follows
\beqn
\Delta K_{bij}^{(2)}=-\frac{2\sqrt 2 }{ g}
 \sum_{l=1}^4   \sum_{k=1}^2  
(\beta_{bl}D_{b1i}+ \alpha_{bl}^* D_{b2i})
(\beta_{bl}D_{b1k}+ \alpha_{bl}^* D_{b2k})
(\alpha_{bj}^*D_{b1k}^*-\gamma_{bj}^* D_{b2k}^*)\nonumber\\
\frac{m_b m_{\chi_l^0}}{16\pi^2} f(m_b^2, m_{\chi_l^0}^2, m_{\tilde b_k}^2)
\label{75}
\eeqn
\beqn
\Delta M_{bij}^{(2)}=-\frac{2\sqrt 2 }{ g}
 \sum_{l=1}^4   \sum_{k=1}^2  
(\alpha_{bl}D_{b1i}- \gamma_{bl}D_{b2i})
(\alpha_{bl}D_{b1k}- \gamma_{bl} D_{b2k})
(\beta_{bj}^*D_{b1k}^*+\alpha_{bj} D_{b2k}^*)\nonumber\\
\frac{m_b m_{\chi_l^0}}{16\pi^2} f(m_b^2, m_{\chi_l^0}^2, m_{\tilde b_k}^2)
\label{76}
\eeqn
Fig.(3c) makes the following contribution
\beqn
\Delta K_{bij}^{(3)}=
\frac{1}{g}
 \sum_{l=1}^3   \sum_{k=1}^2  
[G_{ki} (Y_{l2} + iY_{l3}\cos\beta) 
+ G_{ik}^* (Y_{l2} - iY_{l3}\cos\beta)\nonumber\\ 
+ H_{ki} (Y_{l1} + iY_{l3}\sin\beta)
+ H_{ik}^* (Y_{l1} - iY_{l3}\sin\beta) ]
(C^S_{bl}+iC^P_{bl})
 [\beta_{bj}D_{b1k} +\alpha^*_{bj} D_{b2k}] 
\nonumber\\
\frac{m_b}{16\pi^2} f(m_b^2, m_{H_l}^2, m_{\tilde b_k}^2)
\label{77}
\eeqn
\beqn
\Delta M_{bij}^{(3)}=
\frac{1}{g}
 \sum_{l=1}^3   \sum_{k=1}^2  
[G_{ki} (Y_{l2} + iY_{l3}\cos\beta) 
+ G_{ik}^* (Y_{l2} - iY_{l3}\cos\beta)\nonumber\\ 
+ H_{ki} (Y_{l1} + iY_{l3}\sin\beta)
+ H_{ik}^* (Y_{l1} - iY_{l3}\sin\beta) ]
(C^S_{bl}-iC^P_{bl})
 [\alpha_{bj}D_{b1k} -\gamma_{bj} D_{b2k}] 
\nonumber\\
\frac{m_b }{16\pi^2} f(m_b^2, m_{H_l}^2, m_{\tilde b_k}^2)
\label{78}
\eeqn
Fig.(3d) contributes as follows
\beqn
\Delta K_{bij}^{(4)}=-
 \sum_{l=1}^4   \sum_{k=1}^3  
[Q'_{lj} (Y_{k1}+iY_{k3}\sin\beta) -S_{lj}' (Y_{k2} + iY_{k3}\cos\beta)]
\nonumber\\
(C^S_{bk}+iC^P_{bk})
 (\beta_{bl} D_{b1i}+\alpha^*_{bl} D_{b2i}) 
\frac{m_b m_{\chi_l^0}}{16\pi^2} f(m_b^2, m_{\chi_l^0}^2, m_{H_k}^2)
\label{79}
\eeqn
\beqn
\Delta M_{bij}^{(4)}=-
 \sum_{l=1}^4   \sum_{k=1}^3  
[Q_{jl}^{'*} (Y_{k1}-iY_{k3}\sin\beta) -S_{jl}^{'*} (Y_{k2} - iY_{k3}\cos\beta)]
\nonumber\\
(C^S_{bk}-iC^P_{bk})
 (\alpha_{bl} D_{b1i}-\gamma_{bl} D_{b2i}) 
\frac{m_b m_{\chi_l^0}}{16\pi^2} f(m_b^2, m_{\chi_l^0}^2, m_{H_k}^2)
\label{80}
\eeqn
 where 
 \beqn
 Q'_{ij}= \frac{1}{\sqrt 2} [ X_{3i}^* (X_{2j}^* -\tan\theta_W X_{1j}^*)]
\nonumber\\
 S'_{ij}= \frac{1}{\sqrt 2} [ X_{4j}^* (X_{2i}^* -\tan\theta_W X_{1i}^*)]
\label{81-82}
\eeqn
Fig.(3e) conributes as follows 
\beqn
\Delta K_{bij}^{(5)}=
-\frac{4\sqrt 2 g}{\cos^2\theta_W}
 \sum_{l=1}^4     L^{'''}_{jl} (\frac{1}{3} \sin^2\theta_W) 
 (\beta_{bl} D_{b1i} +\alpha^*_{bl}D_{b2i})
\frac{m_b m_{\chi_l^0}}{16\pi^2} f(m_b^2, m_{Z}^2, m_{\chi_l^0}^2)
\label{83}
\eeqn
\beqn
\Delta M_{bij}^{(5)}=
\frac{4\sqrt 2 g}{\cos^2\theta_W}
 \sum_{l=1}^4    R^{'''}_{jl} 
 (\frac{1}{2}-\frac{1}{3} \sin^2\theta_W) 
 (\alpha_{bl} D_{b1i} -\gamma_{bl}D_{b2i})
\frac{m_b m_{\chi_l^0}}{16\pi^2} f(m_b^2, m_{Z}^2, m_{\chi_l^0}^2)
\label{84}
\eeqn
\beqn
L^{'''}_{ij}= -\frac{1}{2} X_{3i}^* X_{3j} +\frac{1}{2}
X_{4i}^*X_{4j}\nonumber\\
 R^{'''}_{ij}= - L^{'''*}_{ij}  
\label{85-86}
\eeqn
The contribution of Fig.(3f) is
\beqn
\Delta K_{bij}^{(6)}=
 \sum_{l=1}^2  (B^S_{bt}+B^P_{bt})  (U_{l1}D_{b1i}-K_b U_{l2}D_{b2i}) 
  \xi_{jl}^* \cos\beta \nonumber\\ 
  \frac{m_t m_{\chi_l^-}}{16\pi^2} f(m_t^2, m_{\chi_l^-}^2, m_{H^-}^2)
\label{87}
\eeqn
\beqn
\Delta M_{bij}^{(6)}=-
 \sum_{l=1}^2  (B^S_{bt}-B^P_{bt})  (K_t V_{l2}^* D_{b1i})  
  \xi_{jl}^{'*} \sin\beta \nonumber\\ 
  \frac{m_t m_{\chi_l^-}}{16\pi^2} f(m_t^2, m_{\chi_l^-}^2, m_{H^-}^2)
\label{88}
\eeqn
Fig.(3g) contributes as follows 
\beqn
\Delta K_{bij}^{(7)}= 0
\label{89}
\eeqn
\beqn
\Delta M_{bij}^{(7)}= \frac{4g^2}{\sqrt 2}
\sum_{l=1}^2 
R_{lj}^{'*} K_t V_{l2}^* D_{b1i}
\frac{m_t m_{\chi_l^-}}{16\pi^2} f(m_t^2, m_{\chi_l^-}^2, m_{W}^2)
\label{90}
\eeqn
The contribution from Fig.(3h) is 
\beqn
\Delta K_{bij}^{(8)}=-\sum_{k=1}^2 \sum_{l=1}^2
\xi'_{jk}\sin\beta (V_{k1}D_{t1l} -K_t V_{k2} D_{t2l}) 
(\eta_{il}\cos\beta + \eta_{il}^{'*} \sin\beta)\nonumber\\  
 \frac{m_{\chi_l^+}}{16\pi^2} f(m_{\chi_k^+}^2, m_{\tilde t_l}^2, m_{H^-}^2)
\label{91}
\eeqn
\beqn
\Delta M_{bij}^{(8)}=\sum_{k=1}^2 \sum_{l=1}^2
\xi_{jk}\cos\beta (K_b U^*_{k2} D_{t1l}) 
(\eta_{il}\cos\beta + \eta_{il}^{'*} \sin\beta)\nonumber\\  
 \frac{m_{\chi_l^+}}{16\pi^2} f(m_{\chi_k^+}^2, m_{\tilde t_l}^2, m_{H^-}^2)
\label{92}
\eeqn
Finally Fig.(3i) gives
\beqn
\Delta K_{bij}^{(9)}= \sum_{s=1}^4  \sum_{k=1}^2 \sum_{l=1}^3
 [G_{ki} (Y_{l2} + i Y_{l3}\cos\beta) 
 + G_{ik}^* (Y_{l2} -iY_{l3} \cos\beta) 
 + H_{ki} (Y_{l1} + iY_{l3} \sin\beta) \nonumber\\
 + H_{ik}^*  (Y_{l1} - iY_{l3} \sin\beta)]
 (\beta_{bs} D_{b1k} +\alpha_{bs}^* D_{b2k})  
[ Q'_{sj} (Y_{l1}+iY_{l3}\sin\beta) - 
S'_{sj} (Y_{l2}+iY_{l3}\cos\beta)]\nonumber\\
\frac{m_{\chi_s^0}}{16\pi^2} f(m_{\chi_s^0}^2, m_{\tilde b_k}^2, m_{H_l}^2)
\label{93}
\eeqn
\beqn
\Delta M_{bij}^{(9)}= \sum_{s=1}^4  \sum_{k=1}^2 \sum_{l=1}^3
 [G_{ki} (Y_{l2} + i Y_{l3}\cos\beta) 
 + G_{ik}^* (Y_{l2} -iY_{l3} \cos\beta) 
 + H_{ki} (Y_{l1} + iY_{l3} \sin\beta) \nonumber\\
 + H_{ik}^*  (Y_{l1} - iY_{l3} \sin\beta)]
 (\alpha_{bs} D_{b1k} -\gamma_{bs} D_{b2k})  
[ Q^{'*}_{js} (Y_{l1}-iY_{l3}\sin\beta) - 
S^{'*}_{js} (Y_{l2}-iY_{l3}\cos\beta)]\nonumber\\
\frac{m_{\chi_s^0}}{16\pi^2} 
f(m_{\chi_s^0}^2, m_{\tilde b_k}^2, m_{H_l}^2)
\label{93}
\eeqn
The sum of the contribution of  the nine diagrams of Fig.(3) gives
$\Delta K_{bij}$ and  $\Delta M_{bij}$ 
\beqn
\Delta K_{bij} =\sum_{n=1}^9 \Delta K_{bij}^{(n)}\nonumber\\
\Delta M_{bij} =\sum_{n=1}^9 \Delta M_{bij}^{(n)}
\label{95}
\eeqn
We note that the diagram corresponding to Fig.(3c) and Fig.(3i) with 
$Z\leftrightarrow H_l^0$, and the diagram corresponding to
Fig.(3h) with  $W^-\leftrightarrow H^-$ vanish in the
zero external momentum approximation. 
\subsection{Loop corrections to $\bar t \tilde t_i \chi_j^0$ interaction}
Loop corrections to $\bar t \tilde t_i \chi_j^0$ interaction, i.e., 
 $\Delta K_{tij}$ and $\Delta M_{tij}$, 
arise from the nine loops of Fig.(4). We now give the explicit computation
of each loop. Fig.(4a) gives
\beqn
\Delta K_{tij}^{(1)}= -\frac{2\sqrt 2\alpha_s}{3\pi g} 
\sum_{k=1}^2 e^{i\xi_3} D_{t1i}D_{t1k} 
(\alpha_{tj}^* D_{t1k}^*- \gamma_{tj}^* D_{t2k}^*) 
m_t m_{\tilde g}  f(m_{t}^2, m_{\tilde g}^2, m_{\tilde t_k}^2) 
\label{96}
\eeqn
\beqn
\Delta M_{tij}^{(1)}= -\frac{2\sqrt 2\alpha_s}{3\pi g} 
\sum_{k=1}^2 e^{-i\xi_3} D_{t2i}D_{t2k} 
(\beta_{tj}^* D_{t1k}^*+ \alpha_{tj} D_{t2k}^*) 
m_t m_{\tilde g}  f(m_{t}^2, m_{\tilde g}^2, m_{\tilde t_k}^2) 
\label{97}
\eeqn
Fig.(4b) gives
\beqn
\Delta K_{tij}^{(2)}=-\frac{2\sqrt 2 }{ g}
 \sum_{l=1}^4   \sum_{k=1}^2  
(\beta_{tl}D_{t1i}+ \alpha_{tl}^* D_{t2i})
(\beta_{tl}D_{t1k}+ \alpha_{tl}^* D_{t2k})
(\alpha_{tj}^*D_{t1k}^*-\gamma_{tj}^* D_{t2k}^*)\nonumber\\
\frac{m_t m_{\chi_l^0}}{16\pi^2} f(m_t^2, m_{\chi_l^0}^2, m_{\tilde t_k}^2)
\label{98}
\eeqn
\beqn
\Delta M_{tij}^{(2)}=-\frac{2\sqrt 2 }{ g}
 \sum_{l=1}^4   \sum_{k=1}^2  
(\alpha_{tl}D_{t1i}- \gamma_{tl}D_{t2i})
(\alpha_{tl}D_{t1k}- \gamma_{tl} D_{t2k})
(\beta_{tj}^*D_{t1k}^*+\alpha_{tj} D_{t2k}^*)\nonumber\\
\frac{m_t m_{\chi_l^0}}{16\pi^2} f(m_t^2, m_{\chi_l^0}^2, m_{\tilde t_k}^2)
\label{99}
\eeqn
Fig.(4c) makes the following contribution
\beqn
\Delta K_{tij}^{(3)}=
\frac{1}{g}
 \sum_{l=1}^3   \sum_{k=1}^2  
[E_{ki} (Y_{l2} + iY_{l3}\cos\beta) 
+ E_{ik}^* (Y_{l2} - iY_{l3}\cos\beta)\nonumber\\ 
+ F_{ki} (Y_{l1} + iY_{l3}\sin\beta)
+ F_{ik}^* (Y_{l1} - iY_{l3}\sin\beta) ]
(C^S_{tl}+i C^P_{tl}) 
 [\beta_{tj}D_{t1k} +\alpha^*_{tj} D_{t2k}] 
\nonumber\\
\frac{m_t}{16\pi^2} f(m_t^2, m_{H_l}^2, m_{\tilde t_k}^2)
\label{100}
\eeqn
\beqn
\Delta M_{tij}^{(3)}=
\frac{1}{g}
 \sum_{l=1}^3   \sum_{k=1}^2  
[E_{ki} (Y_{l2} + iY_{l3}\cos\beta) 
+ E_{ik}^* (Y_{l2} - iY_{l3}\cos\beta)\nonumber\\ 
+ F_{ki} (Y_{l1} + iY_{l3}\sin\beta)
+ F_{ik}^* (Y_{l1} - iY_{l3}\sin\beta) ]
(C^S_{tl}-iC^P_{tl})
 [\alpha_{tj}D_{t1k} -\gamma_{tj} D_{t2k}] 
\nonumber\\
\frac{m_t}{16\pi^2} f(m_t^2, m_{H_l}^2, m_{\tilde t_k}^2)
\label{101}
\eeqn
 Fig.(4d) gives
\beqn
\Delta K_{tij}^{(4)}=-
 \sum_{l=1}^4   \sum_{k=1}^3  
[Q'_{lj} (Y_{k1}+iY_{k3}\sin\beta) -S_{lj}' (Y_{k2} + iY_{k3}\cos\beta)]
\nonumber\\
(C^S_{tk}+iC^P_{tk})
 (\beta_{tl} D_{t1i}+\alpha_{tl}^* D_{t2i}) 
\frac{m_t m_{\chi_l^0}}{16\pi^2} f(m_t^2, m_{\chi_l^0}^2, m_{H_k}^2)
\label{102}
\eeqn
\beqn
\Delta M_{tij}^{(4)}=-
 \sum_{l=1}^4   \sum_{k=1}^3  
[Q_{jl}^{'*} (Y_{k1}-iY_{k3}\sin\beta) -S_{jl}^{'*} (Y_{k2} - iY_{k3}\cos\beta)]
\nonumber\\
(C^S_{tk}-iC^P_{tk})
 (\alpha_{tl} D_{t1i}-\gamma_{tl} D_{t2i}) 
\frac{m_t m_{\chi_l^0}}{16\pi^2} f(m_t^2, m_{\chi_l^0}^2, m_{H_k}^2)
\label{103}
\eeqn
Fig.(4e) gives
\beqn
\Delta K_{tij}^{(5)}=
\frac{4\sqrt 2 g}{\cos^2\theta_W}
 \sum_{l=1}^4  L^{'''}_{jl} (\frac{2}{3} \sin^2\theta_W) 
 (\beta_{tl} D_{t1i} +\alpha^*_{tl}D_{t2i})
\frac{m_t m_{\chi_l^0}}{16\pi^2} f(m_t^2, m_{Z}^2, m_{\chi_l^0}^2)
\label{104}
\eeqn
\beqn
\Delta M_{tij}^{(5)}=
\frac{-4\sqrt 2 g}{\cos^2\theta_W}
 \sum_{l=1}^4   R^{'''}_{jl} 
 (\frac{1}{2}-\frac{2}{3} \sin^2\theta_W) 
 (\alpha_{tl} D_{t1i} -\gamma_{tl}D_{t2i})
\frac{m_t m_{\chi_l^0}}{16\pi^2} f(m_t^2, m_{Z}^2, m_{\chi_l^0}^2)
\label{105}
\eeqn
Fig.(4f) gives
\beqn
\Delta K_{tij}^{(6)}=
 \sum_{l=1}^2  (B^{S*}_{bt}-B^{P*}_{bt})  (V_{l1}D_{t1i}-K_t V_{l2}D_{t2i}) 
  \xi_{jl}^{'} \sin\beta \nonumber\\
  \frac{m_b m_{\chi_l^+}}{16\pi^2} f(m_b^2, m_{\chi_l^+}^2, m_{H^+}^2)
\label{106}
\eeqn
\beqn
\Delta M_{tij}^{(6)}=-
 \sum_{l=1}^2  (B^{S*}_{bt}+B^{P*}_{bt})  (K_b U_{l2}^* D_{t1i})  
  \xi_{jl} \cos\beta \nonumber\\ 
  \frac{m_b m_{\chi_l^+}}{16\pi^2} f(m_b^2, m_{\chi_l^+}^2, m_{H^+}^2)
\label{107}
\eeqn
Fig.(4g) gives 
\beqn
\Delta K_{tij}^{(7)}= 0
\label{108}
\eeqn
\beqn
\Delta M_{tij}^{(7)}= \frac{4g^2}{\sqrt 2}
\sum_{l=1}^2 
R_{jl}^{'} K_b U_{l2}^* D_{t1i}
\frac{m_b m_{\chi_l^+}}{16\pi^2} f(m_b^2, m_{\chi_l^+}^2, m_{W}^2)
\label{109}
\eeqn
Fig.(4h) makes the following contribution
\beqn
\Delta K_{tij}^{(8)}=-\sum_{k=1}^2 \sum_{l=1}^2
\xi^*_{jk}\cos\beta (U_{k1}D_{b1l} -K_b U_{k2} D_{b2l}) 
(\eta_{li}'\sin\beta + \eta_{li}^{*} \cos\beta)\nonumber\\  
 \frac{m_{\chi_l^-}}{16\pi^2} f(m_{\chi_k^-}^2, m_{\tilde b_l}^2, m_{H^+}^2)
\label{110}
\eeqn
\beqn
\Delta M_{tij}^{(8)}=\sum_{k=1}^2 \sum_{l=1}^2
\xi_{jk}^{'*} \sin\beta (K_t V^*_{k2} D_{b1l}) 
(\eta_{li}'\sin\beta + \eta_{li}^{*} \cos\beta)\nonumber\\  
 \frac{m_{\chi_l^-}}{16\pi^2} f(m_{\chi_k^-}^2, m_{\tilde b_l}^2, m_{H^+}^2)
\label{92}
\eeqn
Finally Fig.(4i) gives
\beqn
\Delta K_{tij}^{(9)}= \sum_{s=1}^4  \sum_{k=1}^2 \sum_{l=1}^3
 [E_{ki} (Y_{l2} + i Y_{l3}\cos\beta) 
 + E_{ik}^* (Y_{l2} -iY_{l3} \cos\beta) 
 + F_{ki} (Y_{l1} + iY_{l3} \sin\beta) \nonumber\\
 + F_{ik}^*  (Y_{l1} - iY_{l3} \sin\beta)]
 (\beta_{ts} D_{t1k} +\alpha_{ts}^* D_{t2k})  
[ Q'_{sj} (Y_{l1}+iY_{l3}\sin\beta) - 
S'_{sj} (Y_{l2}+iY_{l3}\cos\beta)]\nonumber\\
\frac{m_{\chi_s^0}}{16\pi^2} f(m_{\chi_s^0}^2, m_{\tilde t_k}^2, m_{H_l}^2)
\label{112}
\eeqn
\beqn
\Delta M_{tij}^{(9)}= \sum_{s=1}^4  \sum_{k=1}^2 \sum_{l=1}^3
 [E_{ki} (Y_{l2} + i Y_{l3}\cos\beta) 
 + E_{ik}^* (Y_{l2} -iY_{l3} \cos\beta) 
 + F_{ki} (Y_{l1} + iY_{l3} \sin\beta) \nonumber\\
 + F_{ik}^*  (Y_{l1} - iY_{l3} \sin\beta)]
 (\alpha_{ts} D_{t1k} -\gamma_{ts} D_{t2k})  
[ Q^{'*}_{js} (Y_{l1}-iY_{l3}\sin\beta) - 
S^{'*}_{js} (Y_{l2}-iY_{l3}\cos\beta)]\nonumber\\
\frac{m_{\chi_s^0}}{16\pi^2} 
f(m_{\chi_s^0}^2, m_{\tilde t_k}^2, m_{H_l}^2)
\label{113}
\eeqn
The sum of contributions above give $\Delta K_{bij}$ and  $\Delta M_{bij}$ so that 
\beqn
\Delta K_{tij} =\sum_{n=1}^9 \Delta K_{tij}^{(n)}\nonumber\\
\Delta M_{tij} =\sum_{n=1}^9 \Delta M_{tij}^{(n)}
\label{114}
\eeqn
As in the previous analysis,  the contribution from 
   Fig.(4c) with the interchange 
$Z\leftrightarrow  H_l^0$ 
vanishes in the zero external momentum approximation since 
the vertex is proportional to the external momentum. 
Similarly, the contribution from 
Fig.(4h) with the interchange $W^-\leftrightarrow H^-$
vanishes and Fig.(4i) with the interchange $Z\leftrightarrow H_l^0$
vanishes in the zero external momentum approximation for the same reason.
We also note that loops where one  of the internal lines is a gluon line
also vanishes in the zero external momentum approximation since the squark-gluon
interaction gives a vertex of $-ig_s (p+p')^{\mu}$ 
which is of course dependent on the momenta.

\section{Loop corrected squark decays into charginos and neutralinos}
Eqs.(\ref{8}) and (\ref{72}) give the loop corrected effective Lagrangian for 
$\bar q \tilde q_i'\chi_j^{\pm}$ and $\bar q \tilde q_i\chi_j^{0}$ 
interactions. 
Next we use this loop corrected Lagrangian to compute the decay 
widths of the third generation squarks into charginos and neutralinos.
Specifically we will analyze the following decays 
\beqn
\tilde b_i \rightarrow t+\chi_j^{-}\nonumber\\
\tilde t_i \rightarrow b+\chi_j^{+}\nonumber\\
\tilde b_i \rightarrow b+\chi_j^{0}\nonumber\\
\tilde t_i \rightarrow t+\chi_j^{0}
\eeqn
To make the analysis more compact we begin by writing both 
Eqs.(\ref{8}) and (\ref{72}) in the following form 
\beqn
{\cal {L}}= \bar f(B_{ij}^S+ B_{ij}^P \gamma_S) f_j \tilde q_i + H.c.
\label{115}
\eeqn
 where $f$ takes on the values $(t,b)$ and $f_j$ stands for
 $\chi_j^{\pm}, \chi_j^0$ while $\tilde q_i$ can be
 $\tilde b_i, \tilde t_i$. The decay width
 $\Gamma (\tilde q_i \rightarrow f_jf)$ is given by 
  \beqn
 \Gamma(\tilde q_i\rightarrow f_jf)= \frac{1}{4\pi m_i^3}
 [(m_j^2+m_f^2-m_i^2)^2 -4m_j^2 m_f^2]^{\frac{1}{2}}\nonumber\\
 \{\frac{1}{2} (|B_{ij}^S|^2+ |B_{ij}^P|^2)(m_i^2-m_j^2-m_f^2) 
 -\frac{1}{2}  (|B_{ij}^S|^2-|B_{ij}^P|^2)2m_jm_f\}
 \label{116}
 \eeqn
 The co-efficients $B_{ij}^S$ and $B_{ij}^P$ contain the loop corrections
 and depend on the CP phases.Thus, for example, the process 
 $\tilde b_i\rightarrow \chi_j^- +t$ gives the co-efficients
 \beqn
 B_{ij}^S=\frac{g}{2}[R_{bij}+L_{bij}+ \Delta R_{bij} + \Delta
 L_{bij}]\nonumber\\
 B_{ij}^P=\frac{g}{2}[R_{bij}-L_{bij}+ \Delta R_{bij} - \Delta
 L_{bij}]
 \label{117-118}
 \eeqn
  where $R_{bij}$, $L_{bij}$, $\Delta L_{bij}$ and $\Delta R_{bij}$ 
are defined by Eqs.(\ref{2-5}), (\ref{43})  (\ref{44}).
\section{Numerical Analysis and Size of Effects}
In this section we discuss in a quantitative fashion the size of loop
effects on the decay widths of the squarks into chargino and neutralinos.
The analysis of Secs.(2-4) is quite general and valid for the minimal supersymmetric standard
model (MSSM). For the sake of numerical analysis we will limit the parameter space
by working within the framework of the  SUGRA models\cite{sugra}. 
Specifically within the framework of the 
extended mSUGRA model including CP phases, we take as our parameter space at the grand 
unification scale to be the following: the universal scalar mass $m_0$, the universal
gaugino mass $m_{1/2}$, the universal trilinear coupling $|A_0|$, the ratio of the
Higgs vacuum expectation values $\tan\beta =<H_2>/<H_1>$ where $H_2$ gives mass to the
up quarks and $H_1$ gives mass to the down quarks and the leptons. In addition, we take
for CP phases the following: the phase $\theta_{\mu}$ of the Higgs mixing parameter $\mu$
so that $\mu=|\mu| e^{i\theta_{\mu}}$, the phase  $\alpha_{A_0}$ of the trilinear coupling
where $A_0=|A_0| e^{i\alpha_{A_0}}$, and the phases $\xi_i$ (i=1,2,3) of the $SU(3)_C$, 
$SU(2)_L$ and $U(1)_Y$ gauginos, so that $\tilde m_i=|\tilde m_i| e^{i\xi_i}$ (i=1,2,3) 
where $m_i$ (i=1,2,3) are the $SU(3)_C$, $SU(2)_L$ and $U(1)_Y$ gaugino masses.
We note that not all the phases are independent and only certain combinations of them 
appear in the analysis\cite{incancel}. 
In the numerical analysis we compute the loop corrections and also analyze 
their dependence on the phases.

In Fig.\ref{figx1} we give a plot of the decay width of the heavy stop  ($\tilde t_1$)  
into light and heavy 
chargino, $\chi_1^+$ and $\chi_2^+$, i.e., a plot of 
$\Gamma(\tilde t_1\rightarrow b\chi_{1,2}^+)$ 
as a function of $\alpha_{A_0}$. 
The plots are given with the analysis done at the tree level and at the level
of the effective Lagrangian including loop corrections. The analysis shows
that the loop effects can produce a correction of as much as  25\% to the
tree level values. Further, the analysis of Fig.\ref{figx1} shows that the dependence
on $\alpha_{A_0}$  is quite significant and both the tree and the loop
corrections are affected by it. From Fig.\ref{figx1} one finds that the variation with
$\alpha_{A_0}$ in the range $(0,\pi)$ can be as much as 40-50\%. 
In Fig.\ref{figx2} a similar plot is given for the decay width 
$\Gamma(\tilde t_1\rightarrow t\chi_{3,4}^0)$  as a function of $\alpha_{A_0}$.
Here one finds that the loop corrections can be as much as 20\% and 
further that the variations with  $\alpha_{A_0}$ can be as much
as 25-30\%.
The effect of $\alpha_{A_0}$  on the decay width  arises from two sources:
(1) $\alpha_{A_0}$ enters the off diagonal elements of the squark mass$^2$ matrix.
So it affects the squark masses that enter in the decay width. In fact, the modification
of the squark masses due to  $\alpha_{A_0}$ can be large enough that a decay channel may close
or open as $\alpha_{A_0}$
 is varied. This phenomenon will be illustrated explicitly later.
 This type of effect appears both at the tree and at the loop level.
(2) The matrix $D_q$ that diagonalizes the squark matrix is sensitive to variations
of $\alpha_{A_0}$ and this variation again affects both the tree and the loop level 
analysis. Thus at  the tree level  the couplings $R_{qij}$, $L_{qij}$,
$K_{qij}$ and $M_{qij}$  depend  on $\alpha_{A_0}$ and similarly at the loop level the
couplings  $\Delta R_{qij}$, $\Delta L_{qij}$, $\Delta K_{qij}$ and $\Delta M_{qij}$
also depend on $\alpha_{A_0}$.  An important phenomena related to the dependence on 
$\alpha_{A_0}$
is that the effects are strongly dependent on the quark mass. This is so because 
phases enter in the squark mass$^2$ matrix via the off diagonal terms in a prominent way
and these off diagonal terms 
are proportional to the quark mass. Because of this, the sensitivity  of the stop decay
widths to  $\alpha_{A_0}$  is far greater than the sensitivity of the sbottom decay width.
The loop corrections are bigger in the case of the stop decay than for the sbottom case due
to the relative difference of their Yukawa couplings.
For this reason in our numerical analysis we will focus mostly on the effects of phases
on stop decays.

Fig.\ref{figx3} is a repeat of Fig.\ref{figx1} with a plot of the light stop  ($\tilde t_2$) decay width
into charginos, i.e., 
$\Gamma(\tilde t_2\rightarrow b\chi_{1,2}^+)$  as a function of 
$\alpha_{A_0}$. Here one finds that while the loop corrections are 
comparable to the case of Fig.\ref{figx1}, the variations of the decay width
 is  more strongly dependent on 
$\alpha_{A_0}$ in this part of the parameter space.
Fig.\ref{figx4} gives an analysis similar to that of Fig.\ref{figx2} where
plots are given for the decay width 
$\Gamma(\tilde t_2\rightarrow t\chi_{3,4}^0)$  as a function of  
$\alpha_{A_0}$. Here  one finds that the loop corrections can be as much as 30\%.
Further, one finds that the variations with  $\alpha_{A_0}$ are now
much stronger than in the case of Fig.\ref{figx2}. Thus the effect of $\alpha_{A_0}$ 
is  large enough that for values of $\alpha_{A_0}\ge 1.3$ (radian) the 
decays  into  $\chi_3^0, \chi_4^0$  are
closed.  The reason for this is purely kinematical, in that the mass of 
$\tilde t_2$ is strongly dependent on $\alpha_{A_0}$ and varies strongly
with $\alpha_{A_0}$ and falls below the kinematical limit to allow for the
decay into $\chi_3^0, \chi_4^0$ for values of $\alpha_{A_0}\ge 1.3$.
In Fig.\ref{figx5}  a plot is given of the decay width 
$\Gamma(\tilde t_1\rightarrow b\chi^+, t\chi^0)$ 
(where we summed over the final states of charginos and neutralinos)
 both at the tree level and 
at the loop level as a function of $\alpha_{A_0}$.

The analysis of Fig.\ref{figx6} is similar to that of Fig.\ref{figx5} except that one is looking at the
decay width of $\tilde t_2$. 
The discontinuities in Fig.\ref{figx6} are kinematical and arise from the closing of 
some of the neutralino final states.
The analysis of Fig.\ref{figx7} is similar to that of Fig.\ref{figx5}
 while
the analysis of Fig.\ref{figx8} is similar to that of Fig.\ref{figx6} except that the plots are made as 
a function of  $\theta_{\mu}$.
 It is interesting to observe that 
 the dependence of the stop widths on $\theta_{\mu}$  in Fig.\ref{figx8}
  appears to be relatively  weaker. 
  This arises because we are summing over the  
 chargino and neutralino final states.
Thus,  for example,  the decay width $\Gamma(\tilde t_1\rightarrow b\chi_1^+)$ increases with
$\theta_{\mu}$ for the parameters of Fig.\ref{figx7} while  $\Gamma(\tilde t_1\rightarrow b\chi_2^+)$
decreases. This results in the 
sum  $\Gamma(\tilde t_1\rightarrow b\chi_1^0,  b\chi_2^+)$  having  only a 
weak dependence on $\theta_{\mu}$. 
 An analysis similar to that of 
Figs.\ref{figx5}-\ref{figx6} but as a function of $\xi_3$ is carried out in
 Figs.\ref{figx9}-\ref{figx10}. One important new feature
of the decay widths here is that the 
 $\xi_3$ dependence of the widths at the tree level is absent while the loop corrected widths
 show a dependence on $\xi_3$.  Here one finds that the loop corrections are typically
 of size 10-15\% while the overall variation with $\xi_3$ can be as large as 20\%. Typically, the loop
 correction to the sbottom decays are small and the dependence on phases is also relatively small. 
 This is exhibited in Fig.\ref{figx11} where the decay width $\Gamma(\tilde b_1\rightarrow t\chi_1^-)$ is  plotted.
 Here one finds that the loop effects are essentially negligible while the variations of the decay 
 width with $\alpha_{A_0}$ is also essentially negligible.
 The reasons for this weak dependence on the phase and the smallness of loop corrections
 have already been explained on an analytical
 basis at the end of the second paragraph of this section. Here we see that the reasoning 
 presented there is borne out by the numerical analysis.   
Thus the largest loop corrections as well
 as the largest variations with phases arise only for the decay of the stops.

The experimental upper limits of the electric dipole moments
are\cite{eedm,nedm,atomic}:~  $|d_e|<4.3\times 10^{-27}$ecm,
 $|d_n|<6.5\times 10^{-26}$ecm and  $|d_{Hg}|<9.0\times 10^{-28}$ecm.
The last constraint for Hg$^{199}$ could be transformed into a constraint
on a specific combination of the chromoelectric dipole moments of u, d and
s quarks\cite{olive}, $C_{Hg}=|d_d^C -d_u^C - 0.012 d_s^C|< 3.0 \times 10^{-26}$cm.
These constraints are satisfied by the cancellation mechanism  in the numerical analysis 
presented above as follows:
In figure 5, 6 and 7 the constraints are satisfied for the inputs  
 $\tan\beta=40$, $m_0=300$ GeV, $m_{1/2}=300$ GeV,
$\xi_1=0.5$ (radian), $\xi_2=.66$ (radian), $\xi_3=.63$
(radian),  $\theta_{\mu}=2.5$ (radian), $\alpha_{A_0}=1.0$ (radian) and $|A_0|=1$.
At this point we have $|d_e|=1.88\times 10^{-27}$ecm, $|d_n|=1.79\times 10^{-27}$ecm
and $C_{Hg}=8.99 \times 10^{-27}$cm.
In figures 8, 9 and 10 they are satisfied for the inputs
$\tan\beta=45$, $m_0=400$ GeV, $m_{1/2}=400$ GeV,
$\xi_1=0.6$ (radian), $\xi_2=.65$ (radian), $\xi_3=.65$
(radian),  $\theta_{\mu}=2.5$ (radian), $\alpha_{A_0}=2.0$ (radian) and $|A_0|=1$.
At this point we have $|d_e|=3.94\times 10^{-27}$ecm, $|d_n|=9.21\times 10^{-27}$ecm
and $C_{Hg}=3.86 \times 10^{-27}$cm.

\section{Conclusion}
In this paper we have analyzed supersymmetric one loop corrections to the
squark-quark-chargino and to the squark-quark-neutralino 
couplings. The  analysis involves the exchange of the gluino, chargino,
neutralino, W, Z, charged Higgs and neutral Higgs. 
With the above analysis the one loop effective Lagrangian for these
interactions was derived. 
The full CP dependence arising from the
soft CP parameters was taken into account  in the analysis.
The effective Lagrangian was then used to obtain the decay of the squarks
into charginos and neutralinos at the one loop order.  A detailed numerical
analysis within extended SUGRA model was then carried out to study the
size of the loop effects and also to study the effect of CP phases on
the decay widths of the squarks into charginos and neutralinos.
The analysis exhibits that the loop corrections to the decays widths
of the stops can be very substantial, i.e., as much as 30\% or more.
Further, the phase dependence of the decay width is found to be very strong producing
a variation of as much as 40-50\% or more. The phases enter in the decay
widths in two ways; in modifying the stop-bottom-chargino, and the 
stop-top-neutralino couplings and in modifying the stop, chargino and
neutralino masses. In some cases the effect of phases is large enough 
to open or close a decay channel. However, a similar analysis for the
decay of the sbottoms shows the effect of loops as well as the effect of
CP phases to be much smaller.  The one loop effective Lagrangian derived
in this paper would be useful in the analysis of squark decays at 
colliders and in connecting experimental data with the underlying
theoretical schemes such as supergravity and string based models. 

\noindent
{\bf Acknowledgments}\\ 
This research was also supported in part by NSF grant PHY-0139967.\\
{\bf Appendix A} \\
For completeness we give below the loop corrections to the Yukawa couplings $\Delta h_b$,
$\delta h_b$ etc that appear in Sec.2. A derivation of these results can be found in
Ref.\cite{Ibrahim:2003ca} and Ref.\cite{Ibrahim:2003tq}.

\beqn\label{b1}
-\Delta h_b = - \sum_{\it i =1}^2
 \sum_{j=1}^2 \frac{2\alpha_s}{3\pi} e^{-i\xi_3}m_{\tilde g}
G_{ij}^* D_{b1i}^{*} D_{b2j}
f(m_{\tilde g}^2,m_{\tilde b_{\it i}}^2,m_{\tilde b_j}^2)\nonumber\\
 -\sum_{i=1}^2\sum_{j=1}^2\sum_{k=1}^2
g^2 E_{ij}^*\{V_{k1}^*D_{t1i}^* -K_t V_{k2}^* D_{t2i}^*\}
(K_b U_{k2}^*D_{t1j})
\frac{m_{\chi_k^+}}{16\pi^2}
f(m_{\chi_k^+}^2,m_{\tilde t_i}^2,m_{\tilde t_j}^2)\nonumber\\
-\sum_{i=1}^2\sum_{j=1}^2\sum_{k=1}^2
g^2 C_{ij}\{V_{i1}^*D_{t1k}^* -K_t V_{i2}^* D_{t2k}^*\}
(K_b U_{j2}^*D_{t1k})
\frac{m_{\chi_i^+}m_{\chi_j^+}}{16\pi^2}
f(m_{\tilde t_k}^2,m_{\chi_i^+}^2,m_{\chi_j^+}^2)\nonumber\\
+\sum_{i=1}^2\sum_{j=1}^2 \sum_{k=1}^4
2G_{ij}^*
\{\alpha_{bk}D_{b1j}-\gamma_{bk}D_{b2j}\}
   \{\beta_{bk}^*D_{b1i}^*+\alpha_{bk}D_{b2i}^*\}
  \frac{m_{\chi_k^0}}{16\pi^2}
f(m_{\chi_k^0}^2,m_{\tilde b_i}^2,m_{\tilde b_j}^2)\nonumber\\
+\sum_{i=1}^4\sum_{j=1}^4 \sum_{k=1}^2
2\Gamma_{ij}
\{\alpha_{bj}D_{b1k}-\gamma_{bj}D_{b2k}\}
   \{\beta_{bi}^*D_{b1k}^*+\alpha_{bi}D_{b2k}^*\}
  \frac{m_{\chi_i^0} m_{\chi_j^0}}{16\pi^2}
f(m_{\tilde b_k}^2, m_{\chi_i^0}^2, m_{\chi_j^0}^2 )
\eeqn
\beqn
\frac{C_{ij}}{\sqrt 2}= -\frac{g}{\sin\beta}
[ \frac{m_{\chi_i^+}}{2M_W} \delta_{ij}
-Q^*_{ij} \cos\beta - R^*_{ij} ]
\eeqn
\beqn
\frac{\Gamma_{ij}}{\sqrt 2}=
-\frac{g}{2\sin\beta} [ \frac{m_{\chi_i^0}}{2M_W} \delta_{ij}
-Q^{''*}_{ij} \cos\beta
- R^{''*}_{ij}]
\eeqn
\beqn
R_{ij}=\frac{1}{2M_W}[\tilde m_2^* U_{i1} V_{j1} + \mu^* U_{i2} V_{j2}]
\nonumber\\
gQ^{''}_{ij}= \frac{1}{2} [ X_{3i}^* (gX_{2j}^* -g' X_{1j}^*) +
(i\leftrightarrow j) ]\nonumber\\
R_{ij}^{''}= \frac{1}{2M_W} [ \tilde m_1^* X_{1i}^*X_{1j}^*
+ \tilde m_2^* X_{2i}^*X_{2j}^*
-\mu^* (X_{3i}^*X_{4j}^* + X_{4i}^*X_{3j}^*) ]
\eeqn
\beqn
-\delta h_b = - \sum_{\it i =1}^2 \sum_{j=1}^2 \frac{2\alpha_s}{3\pi} e^{-i\xi_3}m_{\tilde g}
H_{ji} D_{b1i}^{*} D_{b2j}
f(m_{\tilde g}^2,m_{\tilde b_{\it i}}^2,m_{\tilde b_j}^2)\nonumber\\
 -\sum_{i=1}^2\sum_{j=1}^2\sum_{k=1}^2
g^2 F_{ji}\{V_{k1}^*D_{t1i}^* -K_t V_{k2}^* D_{t2i}^*\}
(K_b U_{k2}^*D_{t1j})
\frac{m_{\chi_k^+}}{16\pi^2}
f(m_{\chi_k^+}^2,m_{\tilde t_i}^2,m_{\tilde t_j}^2)\nonumber\\
+\sum_{i=1}^2\sum_{j=1}^2 \sum_{k=1}^4
2H_{ji}
\{\alpha_{bk}D_{b1j}-\gamma_{bk}D_{b2j}\}
   \{\beta_{bk}^*D_{b1i}^*+\alpha_{bk}D_{b2i}^*\}
  \frac{m_{\chi_k^0}}{16\pi^2}
f(m_{\chi_k^0}^2,m_{\tilde b_i}^2,m_{\tilde b_j}^2)
\eeqn
\beqn\label{b1}
-\Delta h_t = - \sum_{\it i =1}^2
 \sum_{j=1}^2 \frac{2\alpha_s}{3\pi} e^{-i\xi_3}m_{\tilde g}
F_{ij}^* D_{t1i}^{*} D_{t2j}
f(m_{\tilde g}^2,m_{\tilde t_{\it i}}^2,m_{\tilde t_j}^2)\nonumber\\
 -\sum_{i=1}^2\sum_{j=1}^2\sum_{k=1}^2
g^2 H_{ij}^*\{U_{k1}^*D_{b1i}^* -K_b U_{k2}^* D_{b2i}^*\}
(K_t V_{k2}^*D_{b1j})
\frac{m_{\chi_k^+}}{16\pi^2}
f(m_{\chi_k^+}^2,m_{\tilde b_i}^2,m_{\tilde b_j}^2)\nonumber\\
-\sum_{i=1}^2\sum_{j=1}^2\sum_{k=1}^2
g^2 K_{ij}^*\{U_{i1}^*D_{b1k}^* -K_b U_{i2}^* D_{b2k}^*\}
(K_t V_{j2}^*D_{b1k})
\frac{m_{\chi_i^+}m_{\chi_j^+}}{16\pi^2}
f(m_{\tilde b_k}^2,m_{\chi_i^+}^2,m_{\chi_j^+}^2)\nonumber\\
+\sum_{i=1}^2\sum_{j=1}^2 \sum_{k=1}^4
2F_{ij}^*
\{\alpha_{tk}D_{t1j}-\gamma_{tk}D_{t2j}\}
   \{\beta_{tk}^*D_{t1i}^*+\alpha_{tk}D_{t2i}^*\}
  \frac{m_{\chi_k^0}}{16\pi^2}
f(m_{\chi_k^0}^2,m_{\tilde t_i}^2,m_{\tilde t_j}^2)\nonumber\\
+\sum_{i=1}^4\sum_{j=1}^4 \sum_{k=1}^2
2\Delta_{ij}^*
\{\alpha_{tj}D_{t1k}-\gamma_{tj}D_{t2k}\}
   \{\beta_{ti}^*D_{t1k}^*+\alpha_{ti}D_{t2k}^*\}
  \frac{m_{\chi_i^0} m_{\chi_j^0}}{16\pi^2}
f(m_{\tilde t_k}^2, m_{\chi_i^0}^2, m_{\chi_j^0}^2 )
\eeqn
where
\beq
K_{ij}=-\sqrt{2} g Q_{ji}
\eeq
\beq
\Delta_{ij}=-\frac{g}{\sqrt{2}} Q^{''}_{ij}
\eeq
\beqn
-\delta h_t = - \sum_{\it i =1}^2 \sum_{j=1}^2 \frac{2\alpha_s}{3\pi} e^{-i\xi_3}m_{\tilde g}
E_{ji} D_{t1i}^{*} D_{t2j}
f(m_{\tilde g}^2,m_{\tilde t_{\it i}}^2,m_{\tilde t_j}^2)\nonumber\\
 -\sum_{i=1}^2\sum_{j=1}^2\sum_{k=1}^2
g^2 G_{ji}\{U_{k1}^*D_{b1i}^* -K_b U_{k2}^* D_{b2i}^*\}
(K_t V_{k2}^*D_{b1j})
\frac{m_{\chi_k^+}}{16\pi^2}
f(m_{\chi_k^+}^2,m_{\tilde b_i}^2,m_{\tilde b_j}^2)\nonumber\\
+\sum_{i=1}^2\sum_{j=1}^2 \sum_{k=1}^4
2E_{ji}
\{\alpha_{tk}D_{t1j}-\gamma_{tk}D_{t2j}\}
   \{\beta_{tk}^*D_{t1i}^*+\alpha_{tk}D_{t2i}^*\}
  \frac{m_{\chi_k^0}}{16\pi^2}
f(m_{\chi_k^0}^2,m_{\tilde t_i}^2,m_{\tilde t_j}^2)
\eeqn
\beqn\label{g}
-\overline{\Delta h}_b = -\sum_{i=1}^2\sum_{j=1}^2
\frac{2\alpha_s}{3\pi} e^{-i\xi_3} D_{b2j}D_{t1i}^* \eta_{ji}^*
m_{\tilde g} f(m_{\tilde g}^2, m_{\tilde t_i}^2,m_{\tilde b_j}^2)\nonumber\\
+
\sum_{i=1}^2\sum_{j=1}^2\sum_{k=1}^4
2\eta_{ji}^* (\alpha_{bk}D_{b1j}- \gamma_{bk}D_{b2j})
(\beta_{tk}^* D_{t1i}^* +\alpha_{tk}D_{t2i}^*)\nonumber\\
\frac{m_{\chi_k^0}}{16\pi^2}
f(m_{\chi_k^0}^2,m_{\tilde t_i}^2,m_{\tilde b_j}^2)
+\sum_{i=1}^2\sum_{j=1}^2\sum_{k=1}^4 
\sqrt 2 g\xi_{ki} \frac{m_{\chi_k^0}m_{\chi_i^-}}{16\pi^2}
[-K_b U_{i2}^*D_{t1j}
(\beta_{tk}^*D_{t1j}^* +\alpha_{tk}D_{t2j}^*)\nonumber\\
f(m_{\tilde t_j}^2,m_{\chi_i^-}^2,m_{\chi_k^0}^2)
+ (\alpha_{bk}D_{b1j} -\gamma_{bk}D_{b2j})
 (U_{i1}^*D_{b1j}^* -K_{b}U_{i2}^*D_{b2j}^*)
f(m_{\tilde b_j}^2,m_{\chi_i^-}^2,m_{\chi_k^0}^2)]
\eeqn
\beqn\label{l}
-\overline{\delta h}_b = \sum_{i=1}^2\sum_{j=1}^2
\frac{2\alpha_s}{3\pi} e^{-i\xi_3}
D_{b2j} D_{t1i}^* \eta_{ji}' m_{\tilde g}
f(m_{\tilde g}^2,m_{\tilde t_i}^2,m_{\tilde b_j}^2) \nonumber\\
-\sum_{i=1}^2\sum_{j=1}^2\sum_{k=1}^4
2 \eta_{ji}' (\alpha_{bk} D_{b1j} -\gamma_{bk} D_{b2j})
(\beta_{tk}^* D_{t1i}^*+ \alpha_{tk} D_{t2i}^*)
\frac{m_{\chi_k^0}}{16\pi^2}
f(m_{\chi_k^0}^2,m_{\tilde t_i}^2,m_{\tilde b_j}^2)
\eeqn
\beqn\label{m}
-\overline{\Delta h}_t = -\sum_{i=1}^2\sum_{j=1}^2
\frac{2\alpha_s}{3\pi} e^{-i\xi_3}
D_{b1i}^* D_{t2j} \eta_{ij}'^*  m_{\tilde g}
f(m_{\tilde g}^2,m_{\tilde b_i}^2,m_{\tilde t_j}^2) \nonumber\\
+\sum_{i=1}^2\sum_{j=1}^2\sum_{k=1}^4
2 \eta_{ij}^{'*} (\alpha_{tk} D_{t1j} -\gamma_{tk} D_{t2j})
(\beta_{bk}^* D_{b1i}^*+ \alpha_{bk} D_{b2i}^*)
\frac{m_{\chi_k^0}}{16\pi^2}
f(m_{\chi_k^0}^2,m_{\tilde b_i}^2,m_{\tilde t_j}^2)
\nonumber\\
+\sum_{i=1}^2\sum_{j=1}^2\sum_{k=1}^4 \sqrt 2 g \xi_{ki}^{*'}
\frac{m_{\chi_k^0} m_{\chi_i^-}}{16\pi^2}
[-K_t V_{i2}^* D_{b1j} (\beta_{bk}^*D_{b1j}^* +\alpha_{bk}D_{b2j}^*)\nonumber\\
f(m_{\tilde b_j}^2,m_{\chi_i^-}^2,m_{\chi_k^0}^2)
+(\alpha_{tk}D_{t1j} -\gamma_{tk}D_{t2j})(V_{i1}^*D_{t1j}^* -K_tV_{i2}^*D_{t2j}^*)
f(m_{\tilde t_j}^2,m_{\chi_i^-}^2,m_{\chi_k^0}^2)]
\eeqn
\beqn\label{o}
-\overline{\delta h}_t = \sum_{i=1}^2\sum_{j=1}^2
\frac{2\alpha_s}{3\pi} e^{-i\xi_3}
D_{b1i}^* D_{t2j} \eta_{ij} m_{\tilde g}
f(m_{\tilde g}^2,m_{\tilde b_i}^2,m_{\tilde t_j}^2) \nonumber\\
-\sum_{i=1}^2\sum_{j=1}^2\sum_{k=1}^4
2 \eta_{ij} (\alpha_{tk} D_{t1j} -\gamma_{tk} D_{t2j})
(\beta_{bk}^* D_{b1i}^*+ \alpha_{bk} D_{b2i}^*)
\frac{m_{\chi_k^0}}{16\pi^2}
f(m_{\chi_k^0}^2,m_{\tilde b_i}^2,m_{\tilde t_j}^2)
\eeqn

\newpage

\begin{figure}
\hspace*{-0.6in}
\centering
\includegraphics[width=8cm,height=8cm]{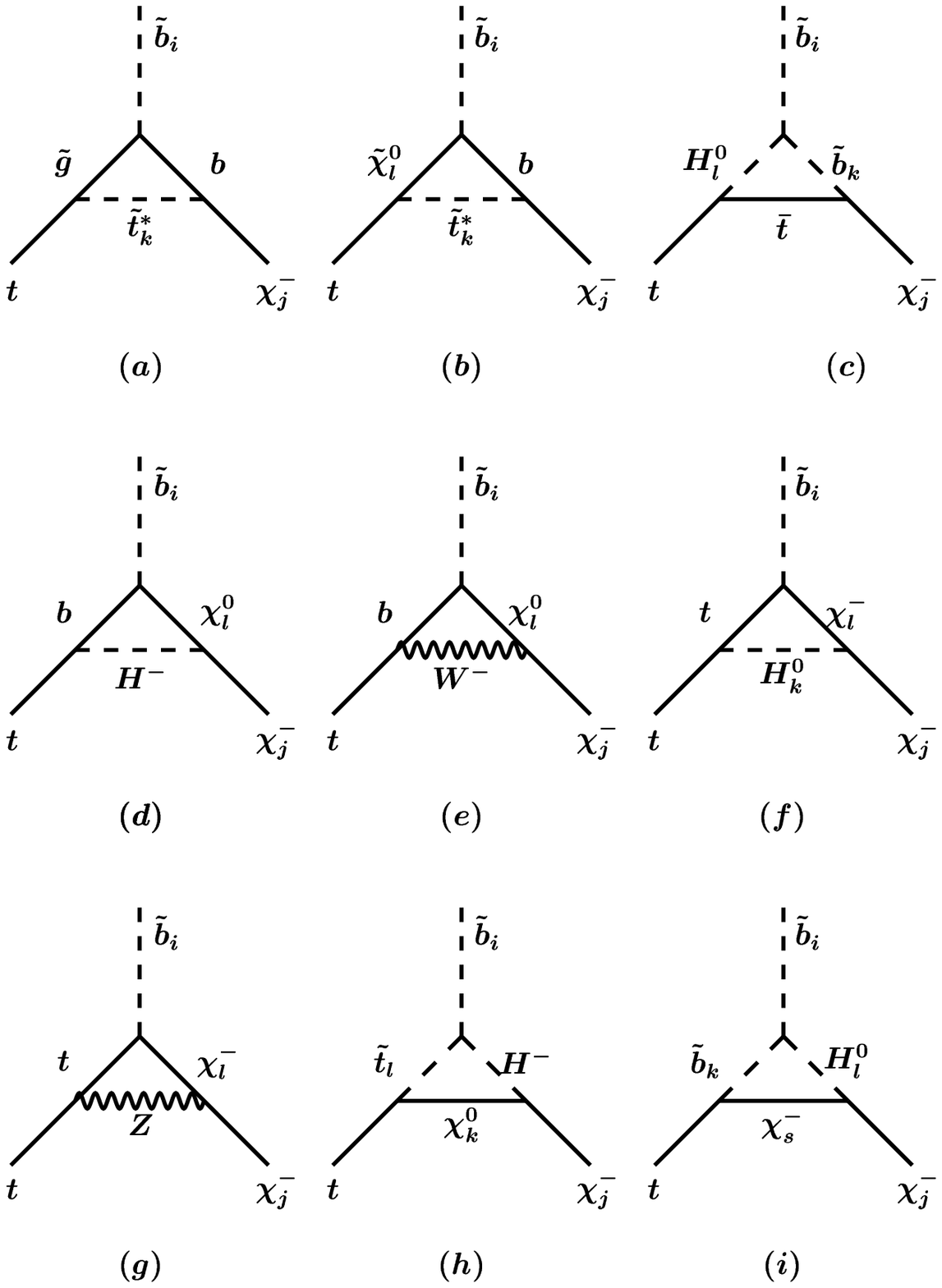}
\caption{List of one loop graphs that contribute to the $\bar{\tilde b_i}t \chi_j^-$ couplings
arising from the exchange of the gluino, charginos, neutralinos, W, Z, charged Higgs  and 
neutral Higgs.}                                                                              		  	                              
\label{fig1A}
\end{figure}

\begin{figure}
\hspace*{-0.6in}
\centering
\includegraphics[width=8cm,height=8cm]{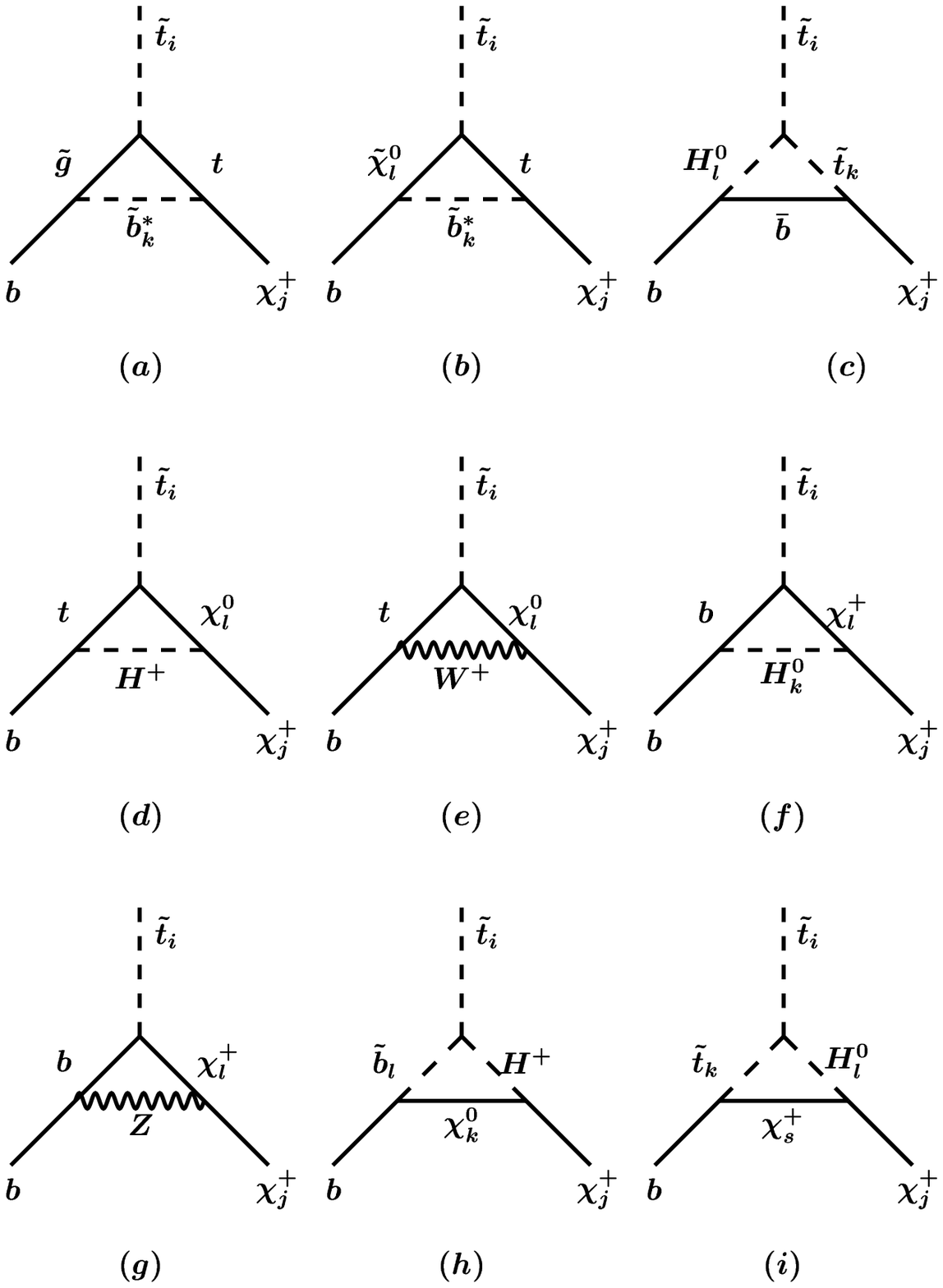}
\caption{List of one loop graphs that contribute to the $\bar{\tilde t_i}b \chi_j^+$ couplings
arising from the exchange of the gluino, charginos, neutralinos, W, Z, charged Higgs  and 
neutral Higgs }                                                                              		  	                              
\label{fig1B}
\end{figure}

\begin{figure}
\hspace*{-0.6in}
\centering
\includegraphics[width=8cm,height=8cm]{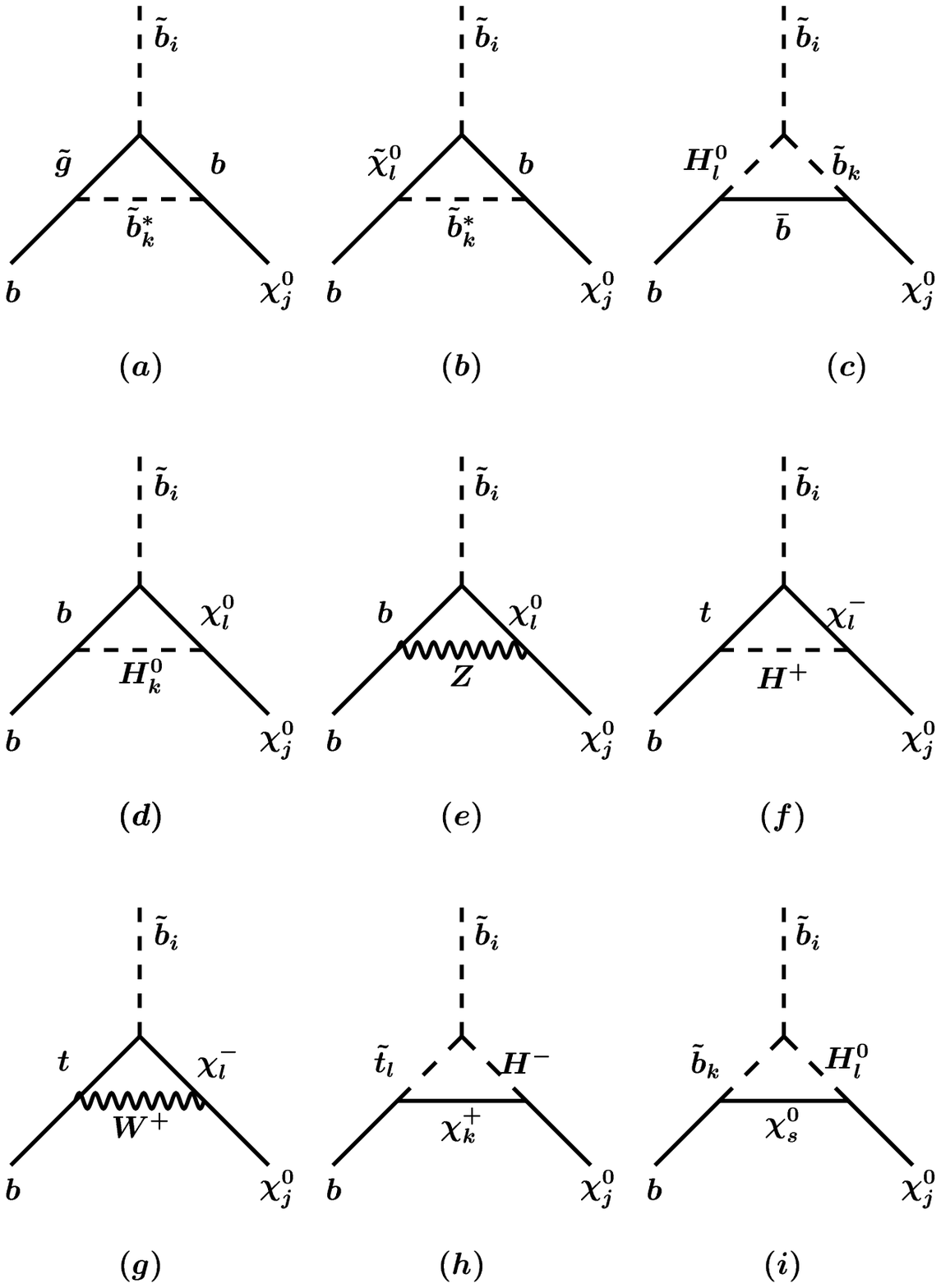}
\caption{List of one loop graphs that contribute to the $\bar{\tilde b_i}b \chi_j^+$ couplings
arising from the exchange of the gluino, charginos, neutralinos, W, Z, charged Higgs  and 
neutral Higgs }                                                                              		  	                              
\label{fig2A}
\end{figure}

\begin{figure}
\hspace*{-0.6in}
\centering
\includegraphics[width=8cm,height=8cm]{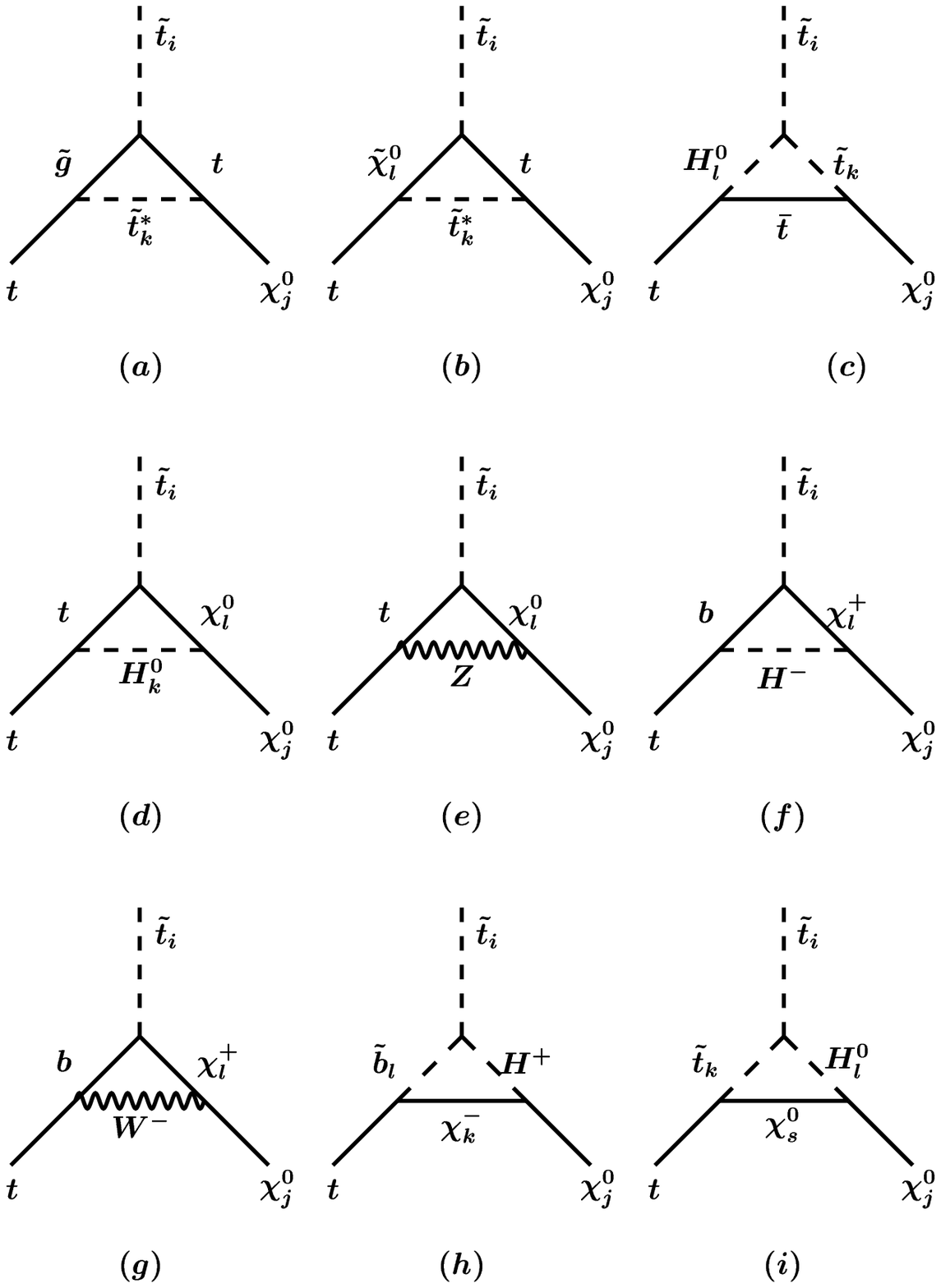}
\caption{List of one loop graphs that contribute to the $\bar{\tilde t_i}t \chi_j^+$ couplings
arising from the exchange of the gluino, charginos, neutralinos, W, Z, charged Higgs  and 
neutral Higgs.}                                                                              		  	                              
\label{fig2B}
\end{figure}

\newpage
\begin{figure}                       
\vspace*{-1.3in}                                 
\subfigure[Plot of the decay width $\Gamma(\tilde t_1\rightarrow b\chi_{1,2}^+)$
as a function of $\alpha_{A_0}$. The solid lines correspond to analysis at the tree 
level while the long-dashed lines include loop corrections. The inputs for the thin lines  
is $\tan\beta=40$, $m_0=300$ GeV, $m_{1/2}=300$ GeV,
$\xi_1=0.5$ (radian), $\xi_2=.66$ (radian), $\xi_3=.63$
(radian),  $\theta_{\mu}=2.5$ (radian), and $|A_0|=1$.  The thick lines are for $\chi_{2}^+$
 decay and the thin lines are for $\chi_{1}^+$ decay.]{
\label{figx1}             
\hspace*{-0.4in}                               
\begin{minipage}[b]{\textwidth}                       
\centering                      
\includegraphics[width=0.9\textwidth, height=0.5\textwidth]{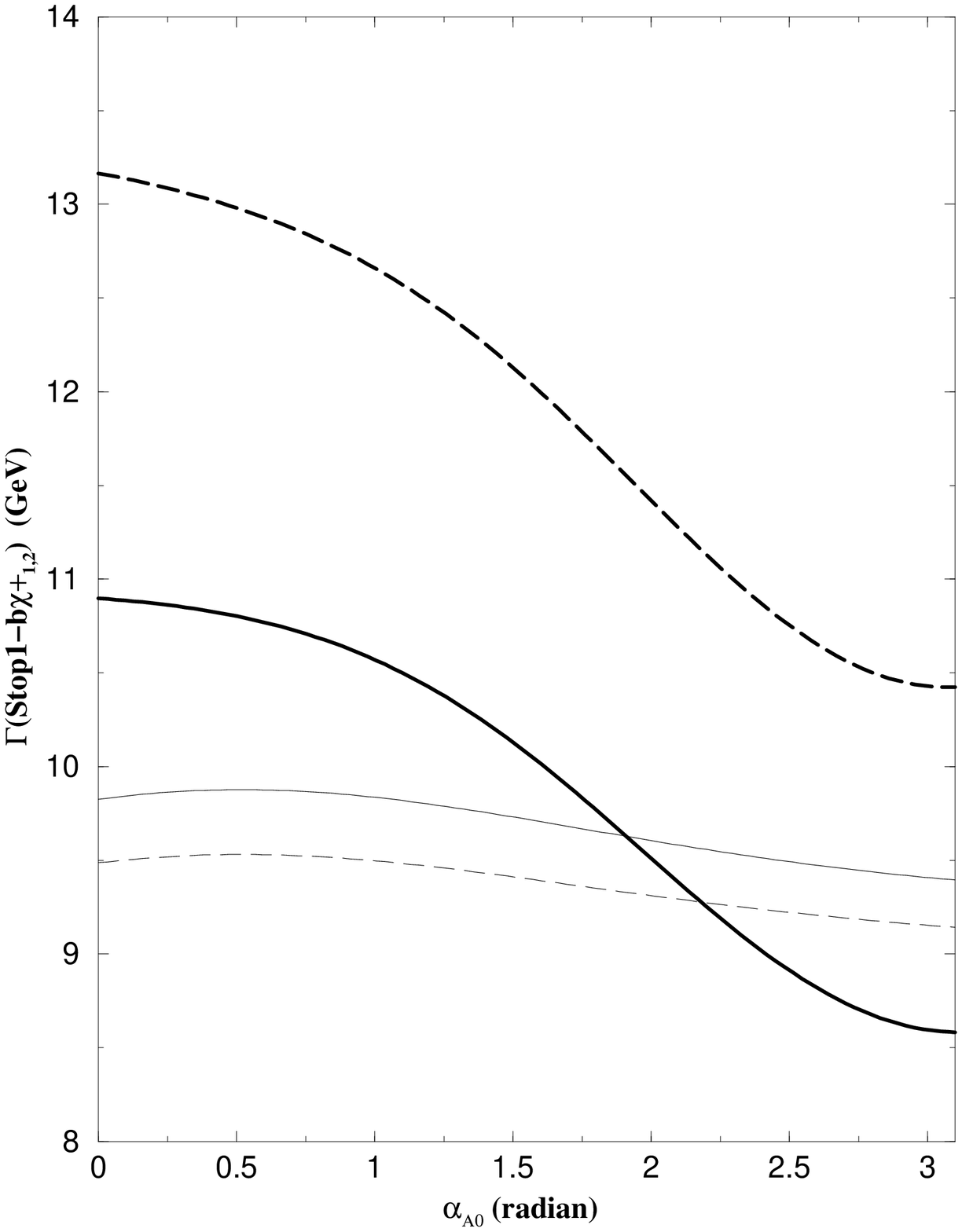} 
\end{minipage}}                       

\subfigure[ Plot of the decay width $\Gamma(\tilde t_1\rightarrow t\chi_{3,4}^0)$
as a function of $\alpha_{A_0}$. The solid lines correspond to analysis at the tree 
level while the long-dashed lines include loop corrections. The inputs for the thin lines  
is $\tan\beta=40$, $m_0=300$ GeV, $m_{1/2}=300$ GeV,
$\xi_1=0.5$ (radian), $\xi_2=.66$ (radian), $\xi_3=.63$
(radian),  $\theta_{\mu}=2.5$ (radian), and $|A_0|=1$.The thick lines are for $\chi_{4}^0$
 decay and the thin lines are for $\chi_{3}^0$ decay.
]{
\label{figx2}             
\hspace*{-0.4in}                               
\begin{minipage}[b]{\textwidth}                       
\centering                      
\includegraphics[width=0.9\textwidth, height=0.5\textwidth]{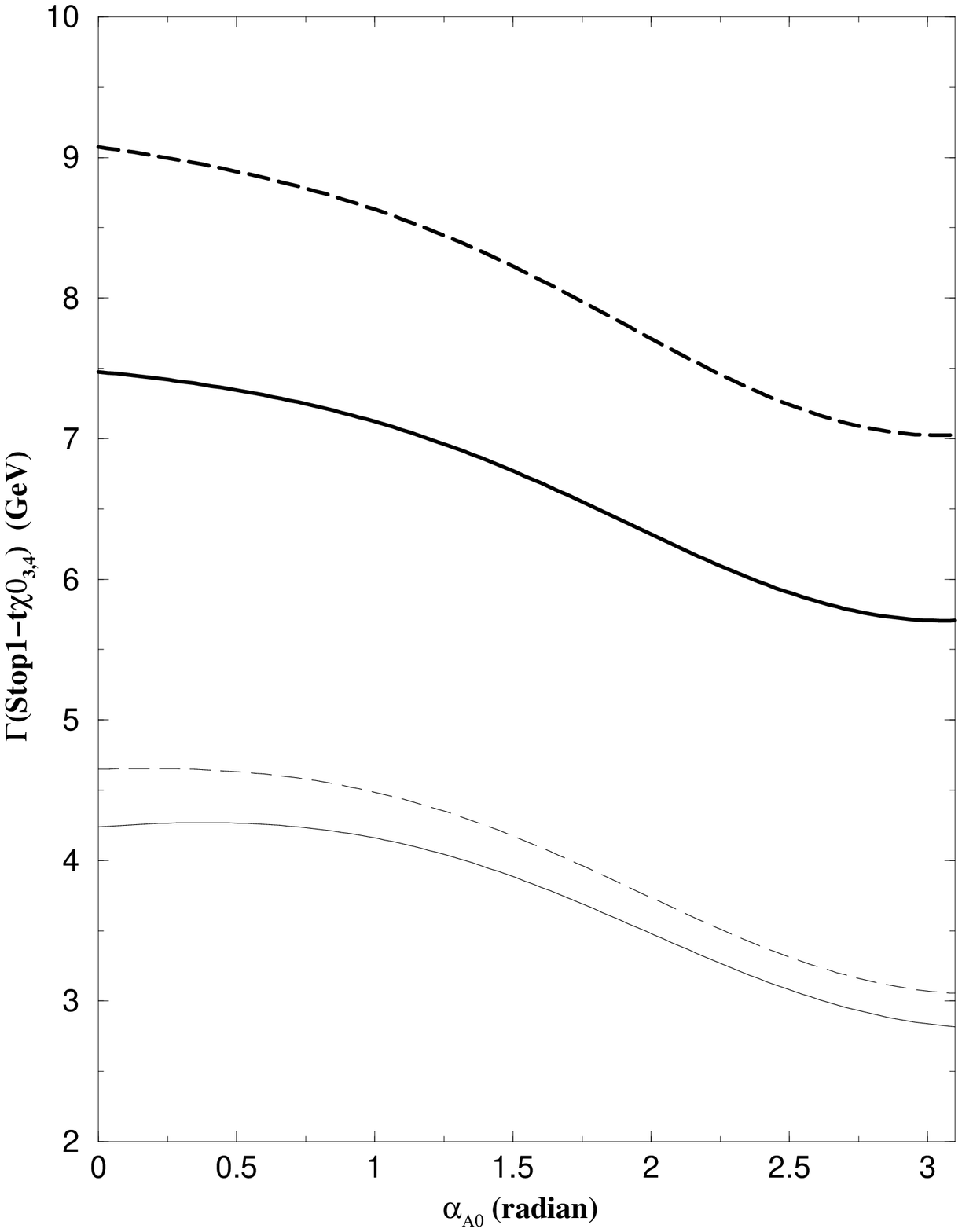} 
\end{minipage}}
\caption{}                       
\label{figx1-2}             
\end{figure}                       


\newpage
\begin{figure}                       
\vspace*{-1.3in}                                 
\subfigure[ Plot of the decay width $\Gamma(\tilde t_2\rightarrow b\chi_{1,2}^+)$
as a function of $\alpha_{A_0}$. The solid lines correspond to analysis at the tree 
level while the long-dashed lines include loop corrections. The inputs for the thin lines  
is $\tan\beta=40$, $m_0=300$ GeV, $m_{1/2}=300$ GeV,
$\xi_1=0.5$ (radian), $\xi_2=.66$ (radian), $\xi_3=.63$
(radian),  $\theta_{\mu}=2.5$ (radian), and $|A_0|=1$.  The thick lines are for $\chi_{2}^+$
 decay and the thin lines are for $\chi_{1}^+$ decay.]{
\label{figx3}             
\hspace*{-0.4in}                               
\begin{minipage}[b]{\textwidth}                       
\centering                      
\includegraphics[width=0.9\textwidth, height=0.5\textwidth]{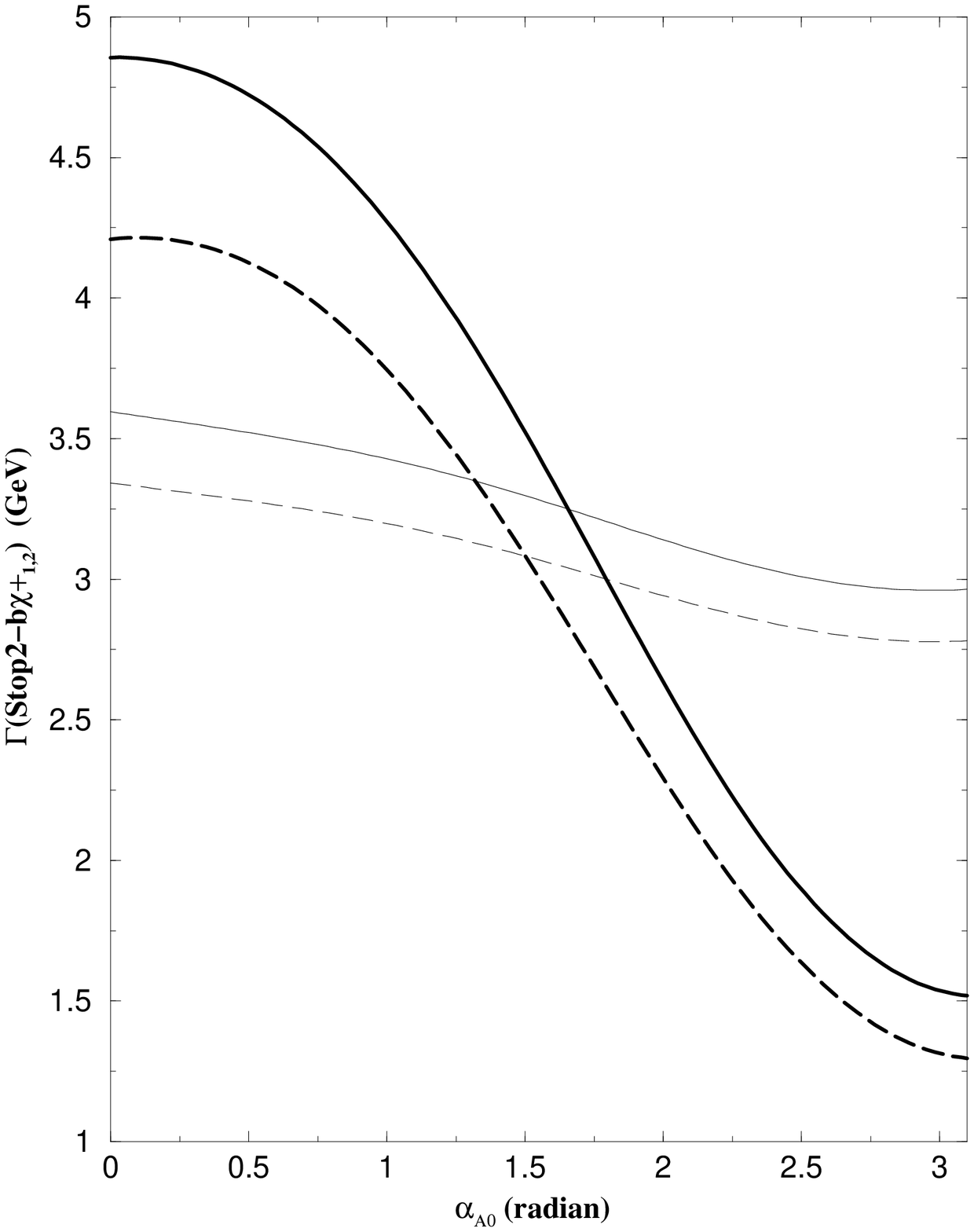} 
\end{minipage}}                       

\subfigure[ Plot of the decay width $\Gamma(\tilde t_2\rightarrow t\chi_{3,4}^0)$
as a function of $\alpha_{A_0}$. The solid lines correspond to analysis at the tree 
level while the long-dashed lines include loop corrections. The inputs for the thin lines  
is $\tan\beta=40$, $m_0=300$ GeV, $m_{1/2}=300$ GeV,
$\xi_1=0.5$ (radian), $\xi_2=.66$ (radian), $\xi_3=.63$
(radian),  $\theta_{\mu}=2.5$ (radian), and $|A_0|=1$. The thick lines are for $\chi_{4}^0$
 decay and the thin lines are for $\chi_{3}^0$ decay.]{
\label{figx4}             
\hspace*{-0.4in}                               
\begin{minipage}[b]{\textwidth}                       
\centering                      
\includegraphics[width=0.9\textwidth, height=0.5\textwidth]{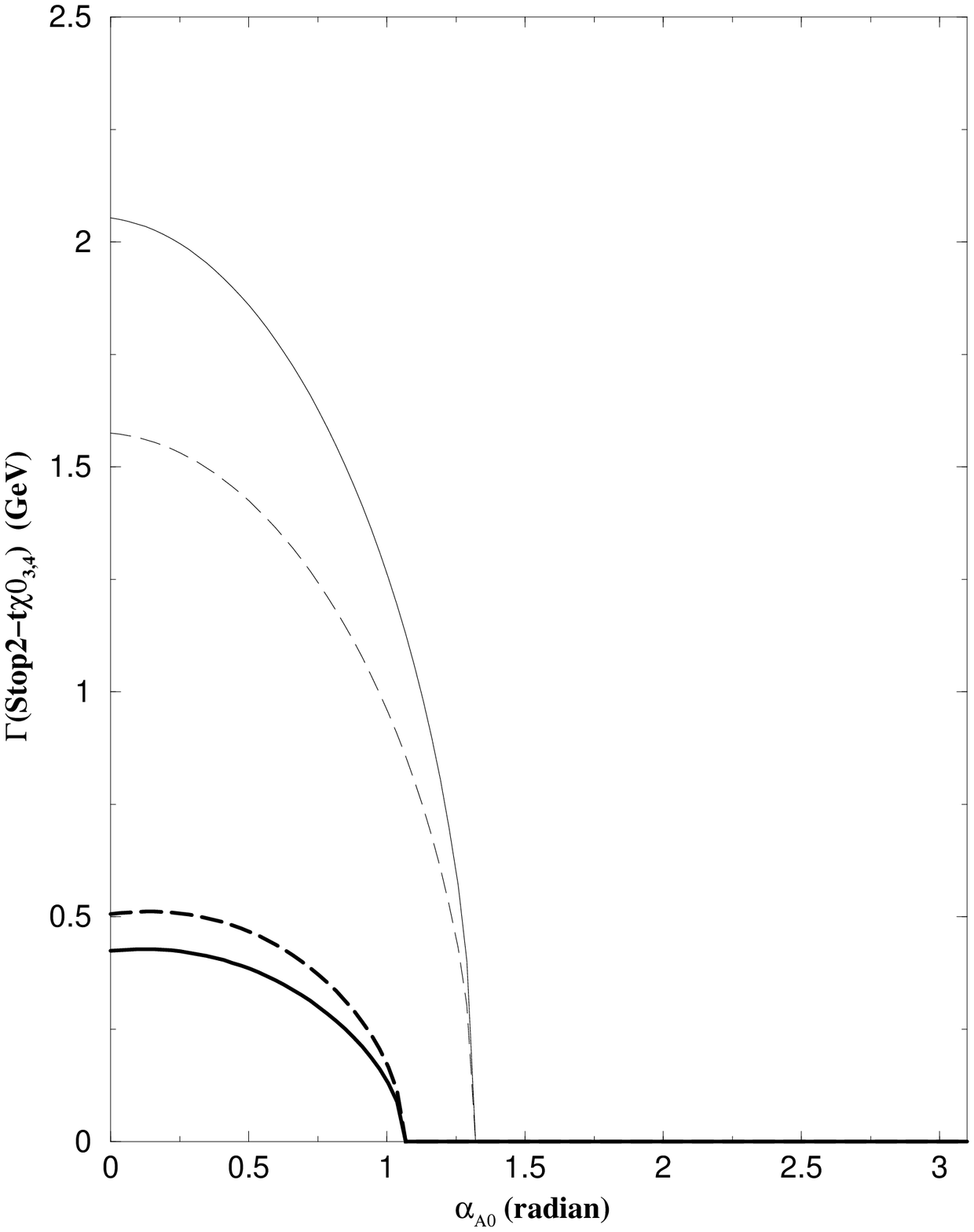} 
\end{minipage}}
\caption{}                       
\label{figx3-4}             
\end{figure}                       


\newpage
\begin{figure}                       
\vspace*{-1.3in}                                 
\subfigure[ Plot of the decay width $\Gamma(\tilde t_1\rightarrow b\chi^+, t\chi^0)$
as a function of $\alpha_{A_0}$. The solid lines correspond to analysis at the tree 
level while the long-dashed lines include loop corrections. The inputs for the thin lines  
is $\tan\beta=40$, $m_0=300$ GeV, $m_{1/2}=300$ GeV,
$\xi_1=0.5$ (radian), $\xi_2=.66$ (radian), $\xi_3=.63$
(radian),  $\theta_{\mu}=2.5$ (radian), and $|A_0|=1$. The thick lines are for the sum over 
the neutralino final states
 and the thin lines are for the sum over the chargino final states. ]{
\label{figx5}             
\hspace*{-0.4in}                               
\begin{minipage}[b]{\textwidth}                       
\centering                      
\includegraphics[width=0.9\textwidth, height=0.5\textwidth]{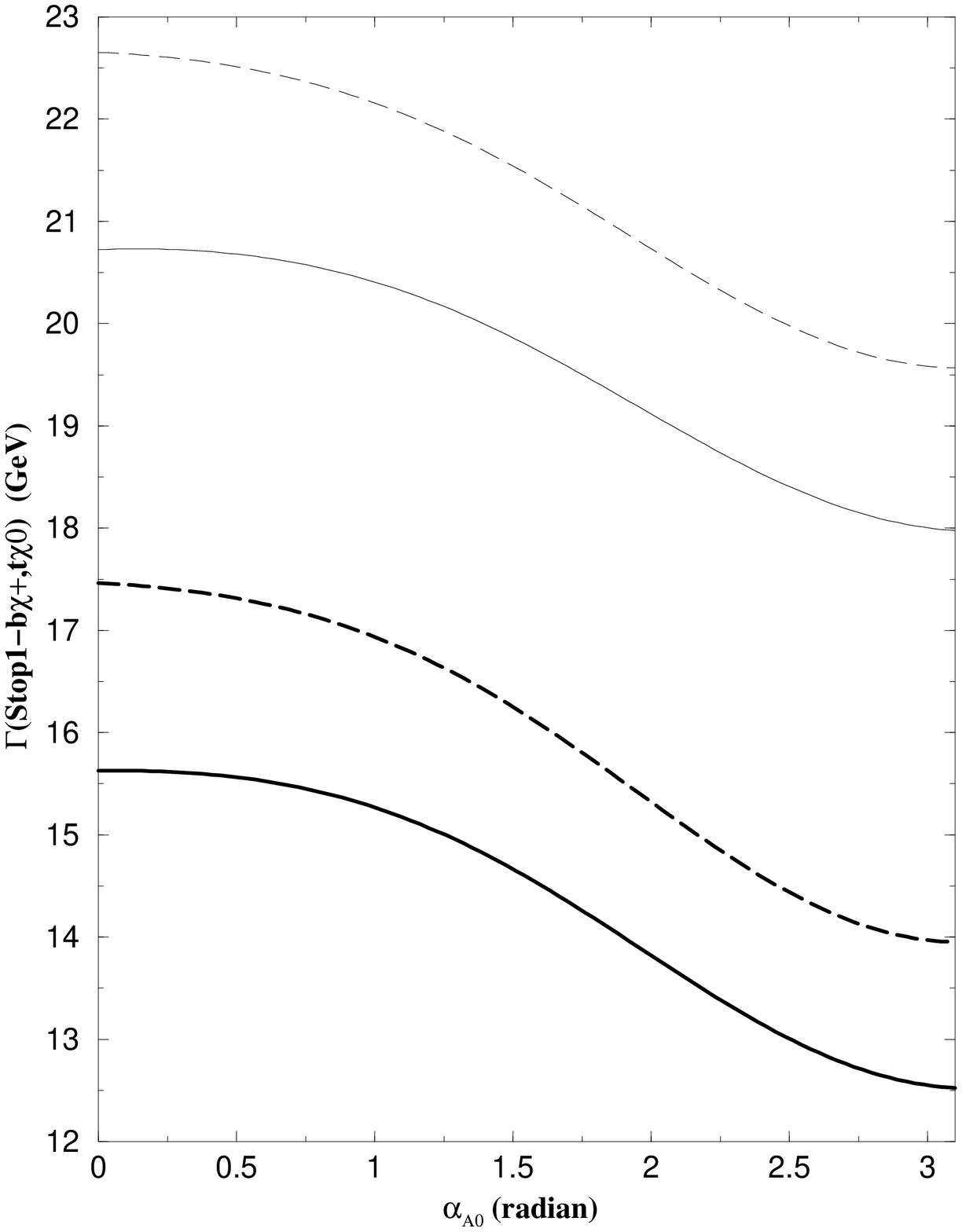} 
\end{minipage}}                       

\subfigure[Plot of the decay width $\Gamma(\tilde t_2\rightarrow b\chi^+, t\chi^0)$
as a function of $\alpha_{A_0}$. The solid lines correspond to analysis at the tree 
level while the long-dashed lines include loop corrections. The inputs for the thin lines  
is $\tan\beta=40$, $m_0=300$ GeV, $m_{1/2}=300$ GeV,
$\xi_1=0.5$ (radian), $\xi_2=.66$ (radian), $\xi_3=.63$
(radian),  $\theta_{\mu}=2.5$ (radian), and $|A_0|=1$.  The thick lines are for the sum over 
the neutralino final states
 and the thin lines are for the sum over the chargino final states.]{
\label{figx6}             
\hspace*{-0.4in}                               
\begin{minipage}[b]{\textwidth}                       
\centering                      
\includegraphics[width=0.9\textwidth, height=0.5\textwidth]{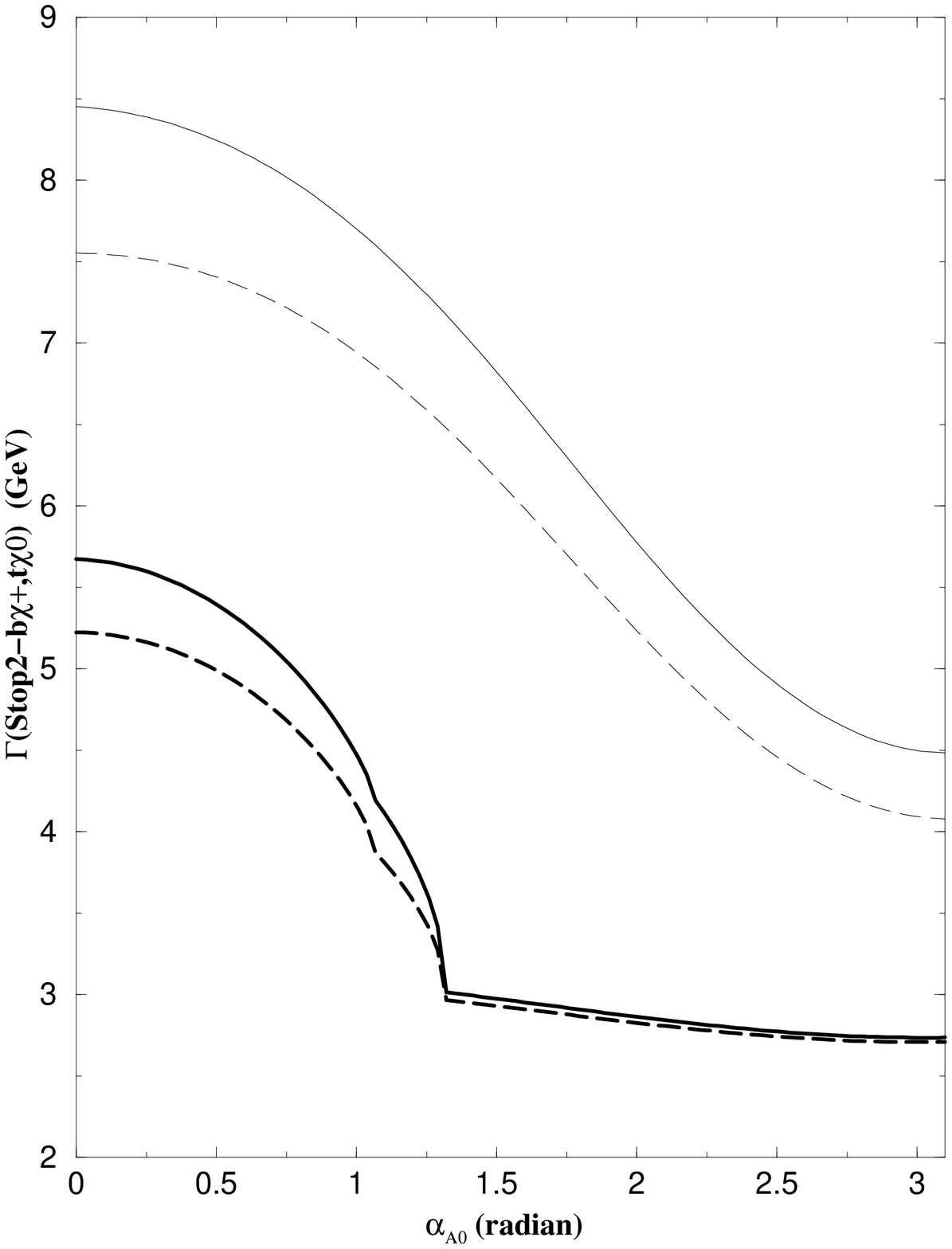} 
\end{minipage}}
\caption{}                       
\label{figx5-6}             
\end{figure}                       


\newpage
\begin{figure}                       
\vspace*{-1.3in}                                 
\subfigure[ Plot of the decay width $\Gamma(\tilde t_1\rightarrow b\chi^+, t\chi^0)$
as a function of $\theta_{\mu}$. The solid lines correspond to analysis at the tree 
level while the long-dashed lines include loop corrections. The inputs for the thin lines  
is $\tan\beta=45$, $m_0=400$ GeV, $m_{1/2}=400$ GeV,
$\xi_1=0.6$ (radian), $\xi_2=.65$ (radian), $\xi_3=.65$
(radian),  $\alpha_{A_0}=2$ (radian), and $|A_0|=1$. The thick lines are for the sum over 
the neutralino final states
 and the thin lines are for the sum over the chargino final states. ]{
\label{figx7}             
\hspace*{-0.4in}                               
\begin{minipage}[b]{\textwidth}                       
\centering                      
\includegraphics[width=0.9\textwidth, height=0.5\textwidth]{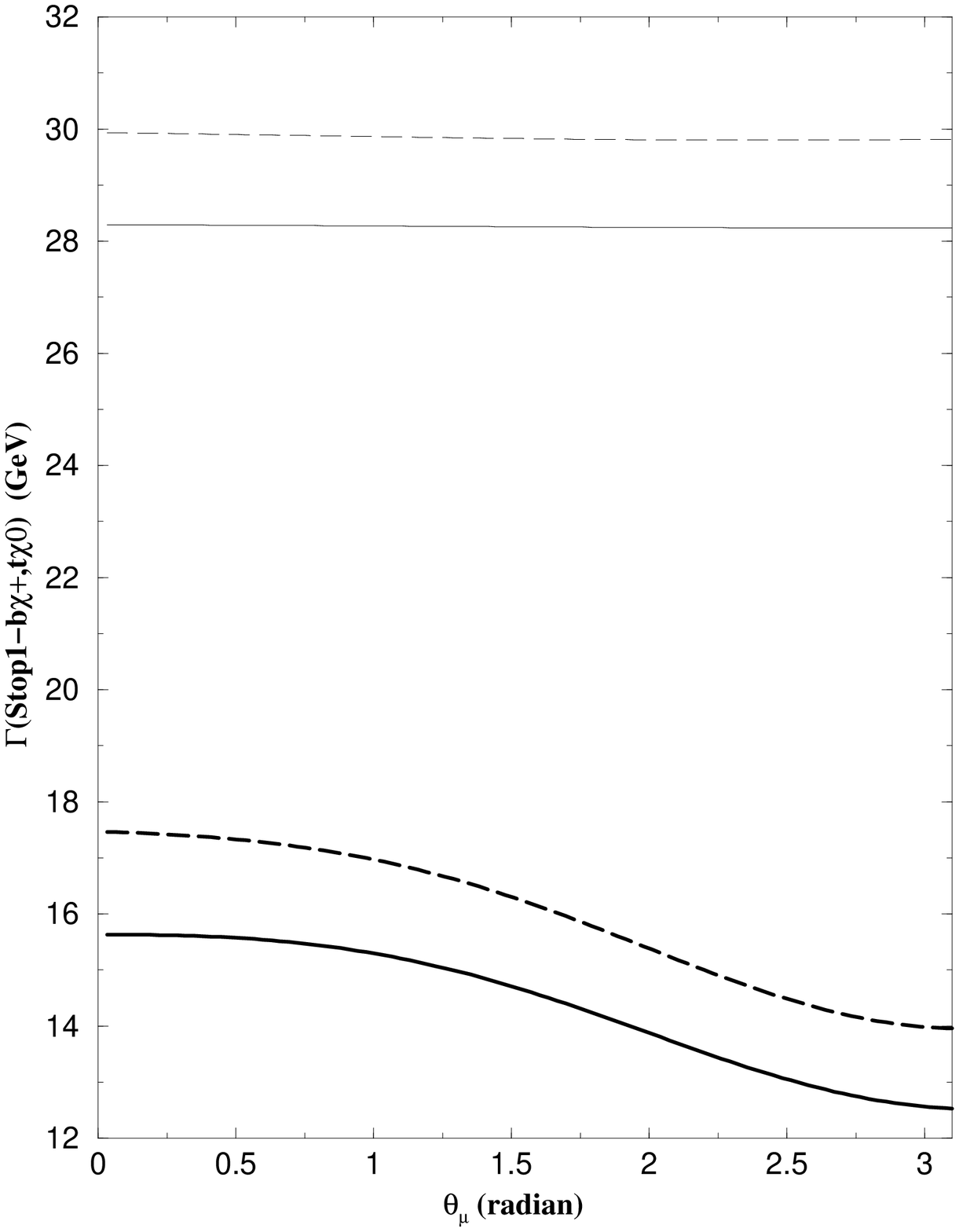} 
\end{minipage}}                       

\subfigure[Plot of the decay width $\Gamma(\tilde t_2\rightarrow b\chi^+, t\chi^0)$
as a function of $\theta_{\mu}$. The solid lines correspond to analysis at the tree 
level while the long-dashed lines include loop corrections. The inputs for the thin lines  
is $\tan\beta=45$, $m_0=400$ GeV, $m_{1/2}=400$ GeV,
$\xi_1=0.6$ (radian), $\xi_2=.65$ (radian), $\xi_3=.65$
(radian),  $\alpha_{A_0}=2$ (radian), and $|A_0|=1$. The thick lines are for the sum over 
the neutralino final states
 and the thin lines are for the sum over the chargino final states.
]{
\label{figx8}             
\hspace*{-0.4in}                               
\begin{minipage}[b]{\textwidth}                       
\centering                      
\includegraphics[width=0.9\textwidth, height=0.5\textwidth]{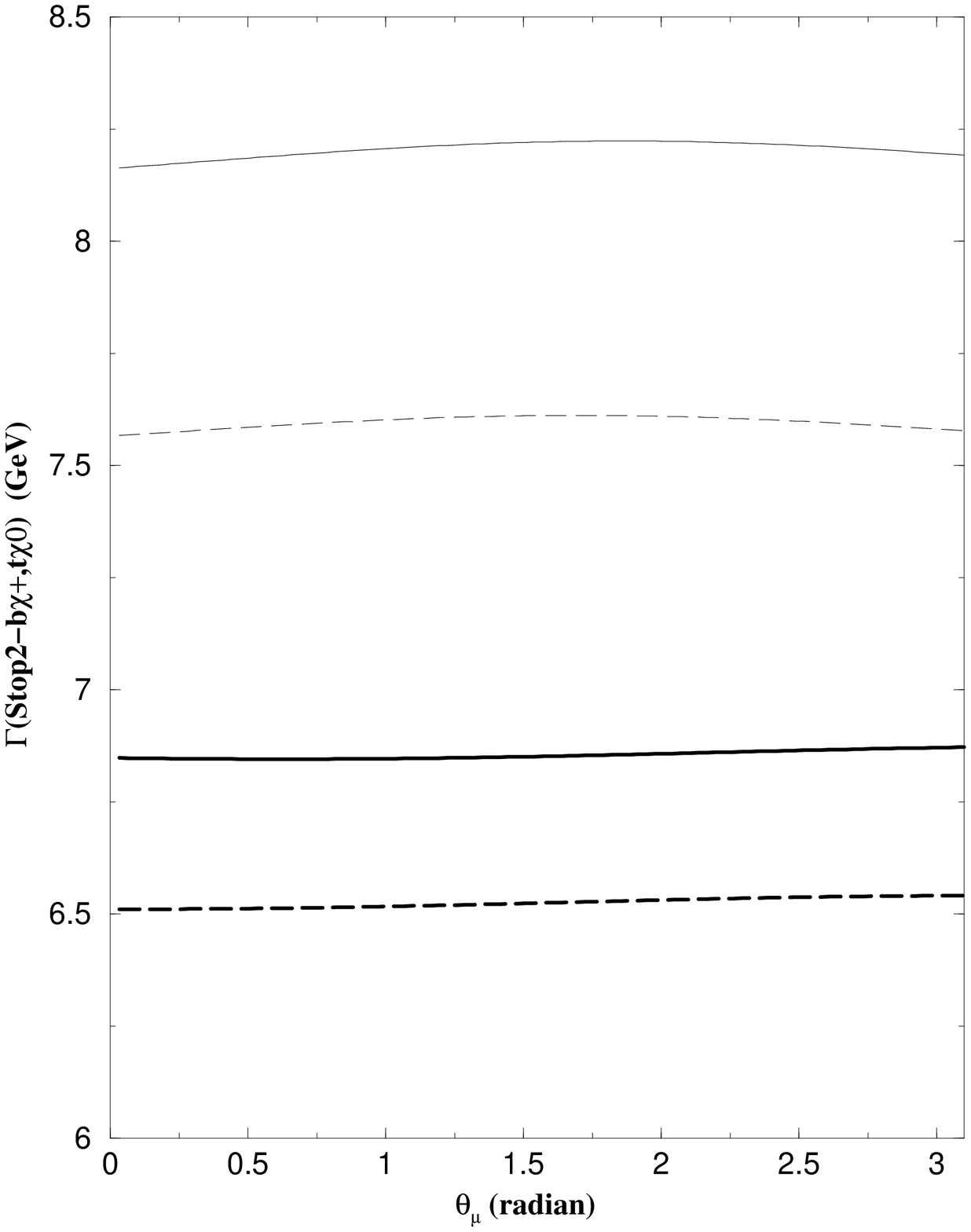} 
\end{minipage}}
\caption{}                       
\label{figx7-8}             
\end{figure}                       


\newpage
\begin{figure}                       
\vspace*{-1.3in}                                 
\subfigure[ Plot of the decay width $\Gamma(\tilde t_1\rightarrow b\chi^+, t\chi^0)$
as a function of $\xi_3$. The solid lines correspond to analysis at the tree 
level while the long-dashed lines include loop corrections. The inputs for the thin lines  
is $\tan\beta=45$, $m_0=400$ GeV, $m_{1/2}=400$ GeV,
$\xi_1=0.6$ (radian), $\xi_2=.65$ (radian), $\theta_{\mu}=2.5$
(radian),  $\alpha_{A_0}=2$ (radian), and $|A_0|=1$. The thick lines are for the sum over 
the neutralino final states
 and the thin lines are for the sum over the chargino final states. ]{
\label{figx9}             
\hspace*{-0.4in}                               
\begin{minipage}[b]{\textwidth}                       
\centering                      
\includegraphics[width=0.9\textwidth, height=0.5\textwidth]{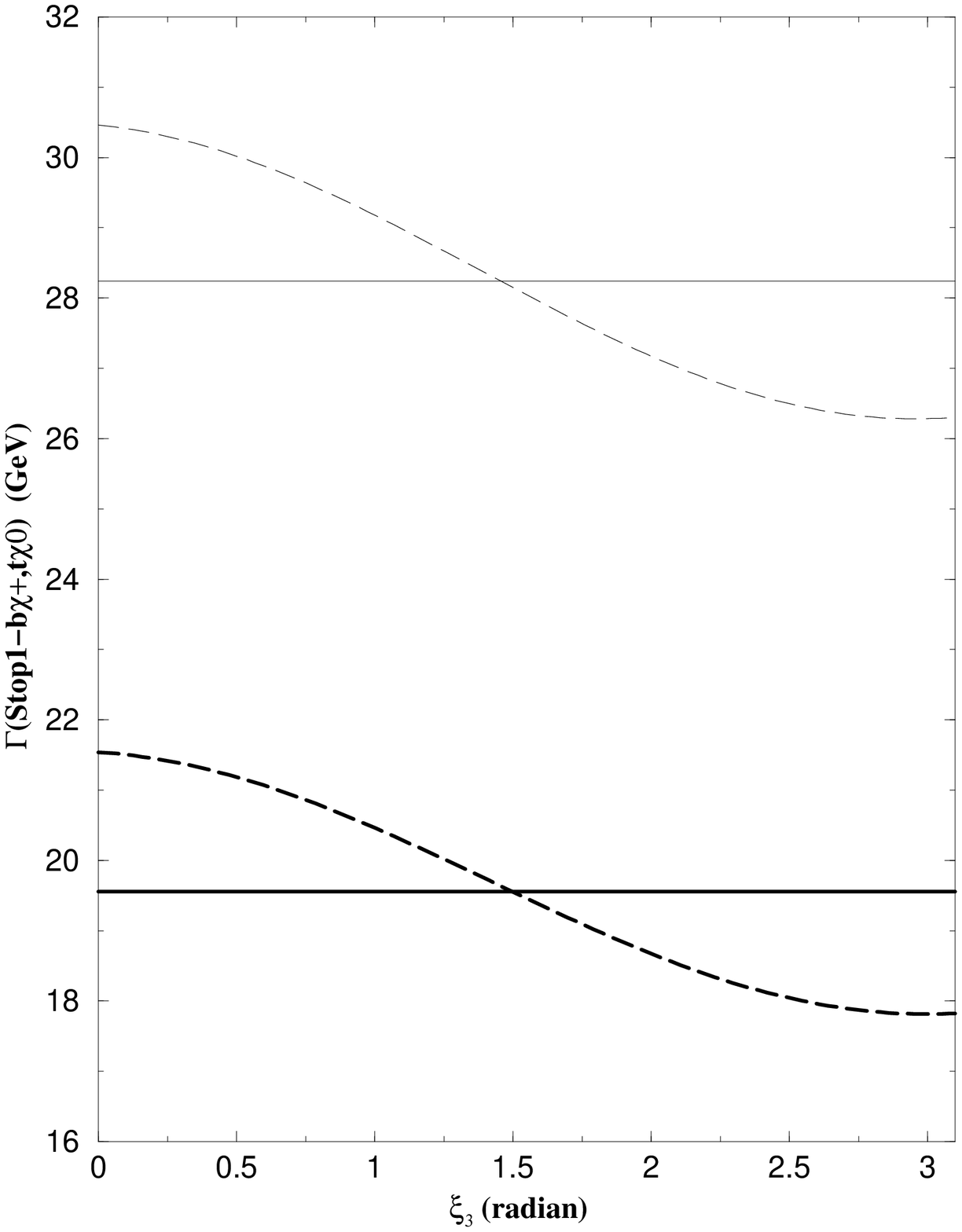} 
\end{minipage}}                       

\subfigure[Plot of the decay width $\Gamma(\tilde t_2\rightarrow b\chi^+, t\chi^0)$
as a function of $\xi_3$. The solid lines correspond to analysis at the tree 
level while the long-dashed lines include loop corrections. The inputs for the thin lines  
is $\tan\beta=45$, $m_0=400$ GeV, $m_{1/2}=400$ GeV,
$\xi_1=0.6$ (radian), $\xi_2=.65$ (radian), $\theta_{\mu}=2.5$
(radian),  $\alpha_{A_0}=2$ (radian), and $|A_0|=1$. The thick lines are for the sum over 
the neutralino final states
 and the thin lines are for the sum over the chargino final states.]{
\label{figx10}             
\hspace*{-0.4in}                               
\begin{minipage}[b]{\textwidth}                       
\centering                      
\includegraphics[width=0.9\textwidth, height=0.5\textwidth]{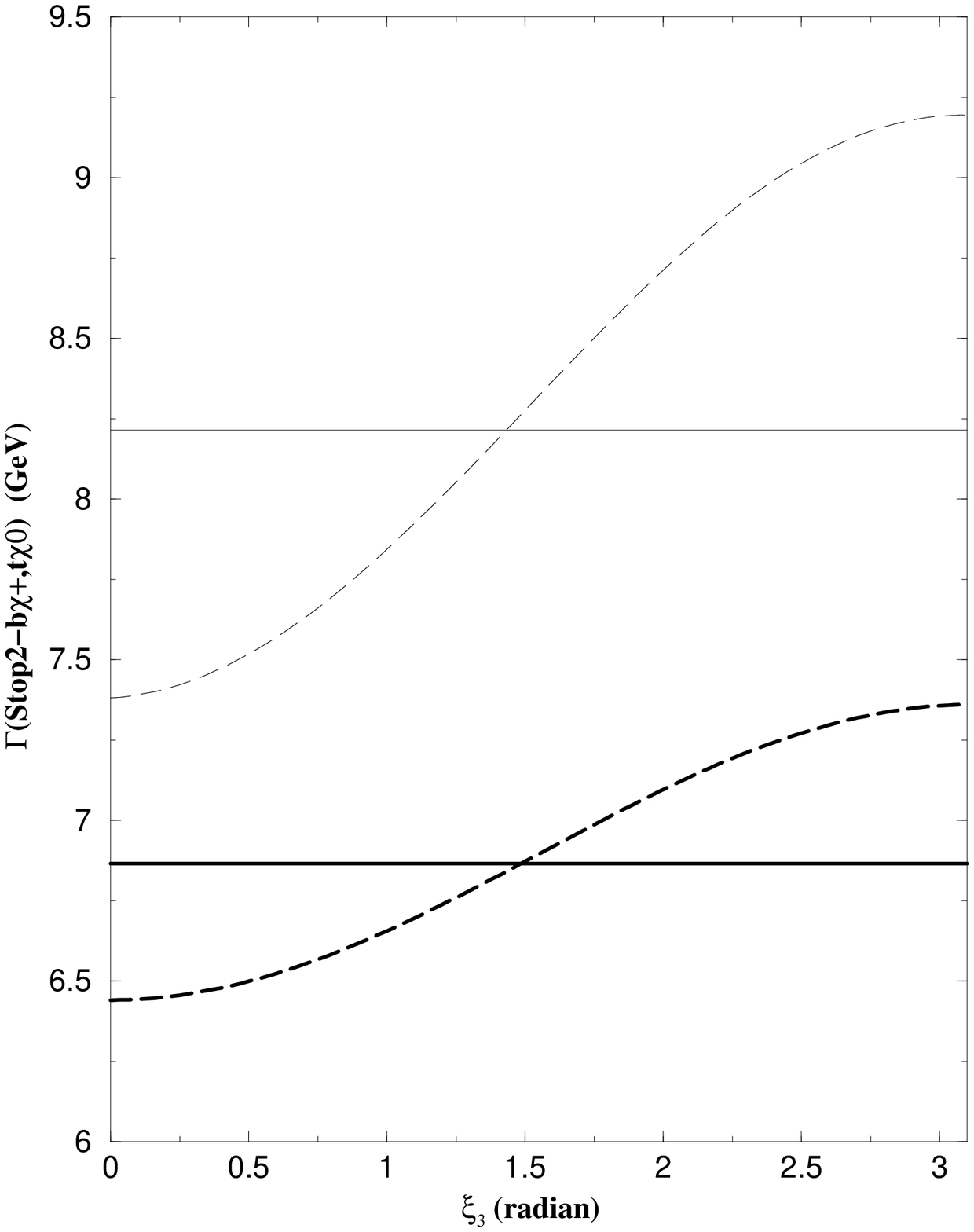} 
\end{minipage}}
\caption{}                       
\label{figx9-10}             
\end{figure}                       


\newpage
\begin{figure}
\hspace*{-1.3in}
\centering
\includegraphics[width=8cm,height=8cm]{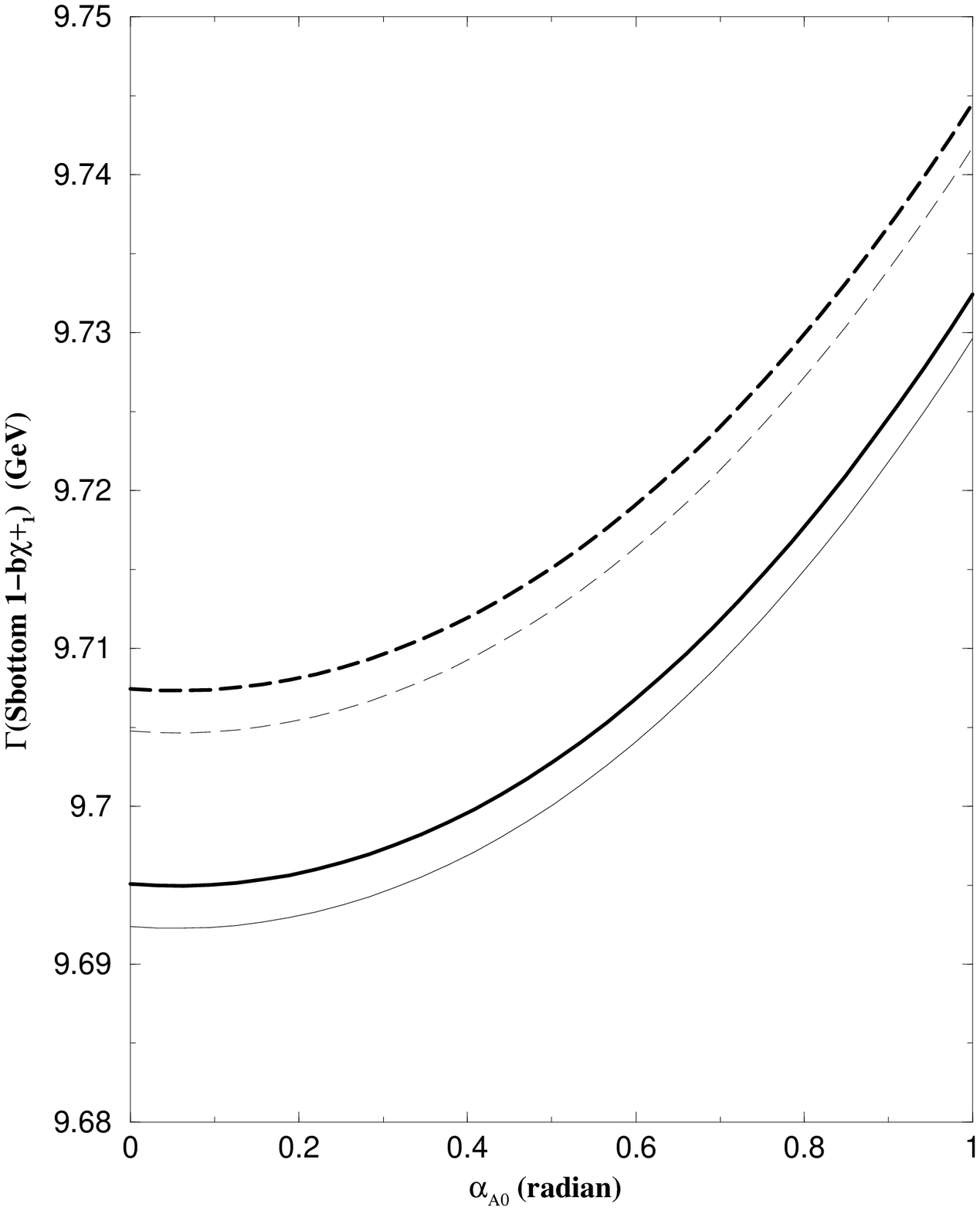}
\caption{ Plot of the decay width $\Gamma(\tilde b_1\rightarrow t\chi^-_1)$
as a function of $\alpha_{A_0}$. The solid lines correspond to analysis at the tree 
level while the long-dashed lines include loop corrections. The inputs for the thin lines  
is $\tan\beta=45$, $m_0=400$ GeV, $m_{1/2}=400$ GeV,
$\xi_1=0.6$ (radian), $\xi_2=.65$ (radian), $\xi_3=.65$ (radian),  $\theta_{\mu}=2.5$
(radian), and $|A_0|=1$.  The input for the thick lines is the  same
as  for the thin lines except that $\xi_2=.5$ (radian).
}                                                                              		  	                              
\label{figx11}
\end{figure}


\begin{thebibliography}{999}

\bibitem{bartl1}
A.~Bartl {\it et al.},
Phys.\ Lett.\ B {\bf 435}, 118 (1998)
[arXiv:hep-ph/9804265];A.~Bartl {\it et al.},
Phys.\ Lett.\ B {\bf 460}, 157 (1999)
[arXiv:hep-ph/9904417];
K.~Hidaka and A.~Bartl,
Phys.\ Lett.\ B {\bf 501}, 78 (2001)
[arXiv:hep-ph/0012021];
C.~Weber, H.~Eberl and W.~Majerotto,
Phys.\ Rev.\ D {\bf 68}, 093011 (2003)
[arXiv:hep-ph/0308146].

\bibitem{bartl2}
A.~Bartl, S.~Hesselbach, K.~Hidaka, T.~Kernreiter and W.~Porod,
Phys.\ Rev.\ D {\bf 70}, 035003 (2004)
[arXiv:hep-ph/0311338].

\bibitem{aoki}
M.~Aoki and N.~Oshimo,
Mod.\ Phys.\ Lett.\ A {\bf 13}, 3225 (1998)
[arXiv:hep-ph/9808217]; W.~M.~Yang and D.~S.~Du,
Phys.\ Rev.\ D {\bf 65}, 115005 (2002)
[arXiv:hep-ph/0202049].

\bibitem{na} 
P.~Nath,
Phys.\ Rev.\ Lett.\  {\bf 66} (1991) 2565
; 
Y. Kizukuri and  N. Oshimo, Phys.Rev.{\bf D46},3025(1992).

 \bibitem{incancel}
 T.~Ibrahim and P.~Nath,
Phys.\ Lett.\ B {\bf 418}, 98 (1998)
;
Phys.\ Rev.\ D {\bf 57}, 478 (1998)
;
Phys.\ Rev.\ D {\bf 58}, 111301 (1998)
 ;
 T. Falk and K Olive, Phys. Lett. {\bf B 439}, 71(1998);
 M. Brhlik, G.J. Good, and G.L. Kane, Phys. Rev. {\bf D59}, 115004
 (1999); A. Bartl, T. Gajdosik, W. Porod, P. Stockinger, and
 H. Stremnitzer,  Phys. Rev. {\bf 60}, 073003(1999);
 S. Pokorski, J. Rosiek and C.A. Savoy, 
 Nucl.Phys. {\bf B570}, 81(2000);
 E.~Accomando, R.~Arnowitt and B.~Dutta,
Phys.\ Rev.\ D {\bf 61}, 115003 (2000);
  U. Chattopadhyay, T. Ibrahim, D.P. Roy, Phys.Rev.D64:013004,2001;
 C.~S.~Huang and W.~Liao,
Phys.\ Rev.\ D {\bf 61}, 116002 (2000);
ibid, Phys.\ Rev.\ D {\bf 62}, 016008 (2000);
 A.Bartl, T. Gajdosik, E.Lunghi, A. Masiero, W. Porod,
H. Stremnitzer and O. Vives, hep-ph/0103324;
 M. Brhlik, L. Everett, G. Kane and J. Lykken, Phys. Rev.
 Lett. {\bf 83}, 2124, 1999; Phys. Rev. {\bf D62}, 035005(2000);
  E. Accomando, R. Arnowitt and B. Datta, 
Phys. Rev. {\bf D61},  075010(2000);
T.~Ibrahim and P.~Nath,
Phys.\ Rev.\ D {\bf 61}, 093004 (2000)
[arXiv:hep-ph/9910553].



\bibitem{olive} 
 T. Falk, K.A. Olive, M. Prospelov, and R. Roiban, Nucl. Phys. 
 {\bf B560}, 3(1999); V.~D.~Barger, T.~Falk, T.~Han, J.~Jiang, T.~Li 
 and T.~Plehn,
Phys.\ Rev.\ D {\bf 64}, 056007 (2001);
S.Abel, S. Khalil, O.Lebedev, Phys. Rev. Lett. {\bf 86}, 5850(2001);
T.~Ibrahim and P.~Nath,
Phys.\ Rev.\ D {\bf 67}, 016005 (2003)
arXiv:hep-ph/0208142.

\bibitem{chang}
D. Chang, W-Y.Keung,and A. Pilaftsis, Phys. Rev. Lett. {\bf 82}, 
900(1999). 

\bibitem{eedm}
E. Commins, et. al., Phys. Rev. {\bf A50}, 2960(1994).

\bibitem{nedm}
P.G. Harris et.al., Phys. Rev. Lett. {\bf 82}, 904(1999).

\bibitem{atomic}
S.~K.~Lamoreaux, J.~P.~Jacobs, B.~R.~Heckel, F.~J.~Raab and E.~N.~Fortson,
Phys.\ Rev.\ Lett.\  {\bf 57}, 3125 (1986).

\bibitem{cphiggsmass}
A. Pilaftsis, Phys. Rev. {\bf D58}, 096010; Phys. Lett.{\bf B435}, 
88(1998);
~A. Pilaftsis and C.E.M. Wagner, Nucl. Phys. {\bf B553}, 3(1999);
~D.A. Demir, Phys. Rev. {\bf D60}, 055006(1999);
~S.~Y.~Choi, M.~Drees and J.~S.~Lee,
Phys.\ Lett.\ B {\bf 481}, 57 (2000);
T.~Ibrahim and P.~Nath,
Phys.\ Rev.\ D {\bf 63}, 035009 (2001)
[arXiv:hep-ph/0008237].
;
T.~Ibrahim,
Phys.\ Rev.\ D {\bf 64}, 035009 (2001);
T.~Ibrahim and P.~Nath,
Phys.\ Rev.\ D {\bf 66}, 015005 (2002);
~S.~W.~Ham, S.~K.~Oh, E.~J.~Yoo, C.~M.~Kim and D.~Son,
arXiv:hep-ph/0205244;
~M.~Boz,
Mod.\ Phys.\ Lett.\ A {\bf 17}, 215 (2002).
;
M.~Carena, J.~R.~Ellis, A.~Pilaftsis and C.~E.~Wagner,
Nucl.\ Phys.\ B {\bf 625}, 345 (2002)
[arXiv:hep-ph/0111245].
;
A.~Dedes and A.~Pilaftsis,
Phys.\ Rev.\ D {\bf 67}, 015012 (2003)
[arXiv:hep-ph/0209306]
;
J.~Ellis, J.~S.~Lee and A.~Pilaftsis,
arXiv:hep-ph/0404167.

\bibitem{Christova:2002sw}
E.~Christova, H.~Eberl, W.~Majerotto and S.~Kraml,
JHEP {\bf 0212}, 021 (2002)
[arXiv:hep-ph/0211063];
E.~Christova, H.~Eberl, W.~Majerotto and S.~Kraml,
Nucl.\ Phys.\ B {\bf 639}, 263 (2002)
[Erratum-ibid.\ B {\bf 647}, 359 (2002)]
[arXiv:hep-ph/0205227].

\bibitem{Ibrahim:2003ca}
T.~Ibrahim and P.~Nath,
Phys.\ Rev.\ D {\bf 67}, 095003 (2003)
[arXiv:hep-ph/0301110].

\bibitem{Ibrahim:2003jm}
T.~Ibrahim and P.~Nath,
Phys.\ Rev.\ D {\bf 68}, 015008 (2003)
[arXiv:hep-ph/0305201].

\bibitem{Ibrahim:2003tq}
T.~Ibrahim and P.~Nath,
Phys.\ Rev.\ D {\bf 69}, 075001 (2004)
[arXiv:hep-ph/0311242].


\bibitem{ibrahim1}
T.~Ibrahim and P.~Nath,
Phys.\ Rev.\ D {\bf 68}, 015008 (2003)
[arXiv:hep-ph/0305201].

\bibitem{ibrahim2}
T.~Ibrahim and P.~Nath,
Phys.\ Rev.\ D {\bf 69}, 075001 (2004)
[arXiv:hep-ph/0311242].

\bibitem{ibrahim3}
T.~Ibrahim, P.~Nath and A.~Psinas,
Phys.\ Rev.\ D {\bf 70}, 035006 (2004)
[arXiv:hep-ph/0404275].

\bibitem{cpdark}
U.~Chattopadhyay, T.~Ibrahim and P.~Nath,
Phys.\ Rev.\ D {\bf 60}, 063505 (1999)
[arXiv:hep-ph/9811362]
;  
 T. Falk, A. Ferstl and K. Olive, Astropart. Phys. {\bf 13}, 301(2000);
P.~Gondolo and K.~Freese,
JHEP {\bf 0207}, 052 (2002)
[arXiv:hep-ph/9908390].


 \bibitem{gomez}
 M.~E.~Gomez, T.~Ibrahim, P.~Nath and S.~Skadhauge,
Phys.\ Rev.\ D {\bf 70}, 035014 (2004)
[arXiv:hep-ph/0404025]
;
arXiv:hep-ph/0410007
;
T.~Nihei and M.~Sasagawa,
arXiv:hep-ph/0404100.

\bibitem{Ibrahim:2002ry}
For a more  complete set of references see, 
T.~Ibrahim and P.~Nath,
``Phases and CP violation in SUSY,''
arXiv:hep-ph/0210251 published in 
P.~Nath and P.~M.~Zerwas,
``Supersymmetry and unification of fundamental interactions. 
Proceedings, 10th International Conference, SUSY'02, Hamburg, Germany, June 17-23,
2002,'' DESY-PROC-2002-02

\bibitem{carena2002}
For a recent review, see, 
M.~Carena and H.~E.~Haber,
Prog.\ Part.\ Nucl.\ Phys.\  {\bf 50}, 63 (2003)
[arXiv:hep-ph/0208209].

\bibitem{sugra}
A.~H.~Chamseddine, R.~Arnowitt and P.~Nath,
Phys.\ Rev.\ Lett.\  {\bf 49}, 970 (1982)
;
 ~R. Barbieri, S. Ferrara and C.A. Savoy, \Journal{\PLB}
{119}{343}{1982}; ~L. Hall, J. Lykken, and S. Weinberg,
\Journal{\PRD}{27}{2359}{1983}
;
P.~Nath, R.~Arnowitt and A.~H.~Chamseddine,
Nucl.\ Phys.\ B {\bf 227}, 121 (1983)
;
For a recent review see, P.~Nath,
``Twenty years of SUGRA,''
arXiv:hep-ph/0307123..


\end{thebibliography}
\end{document}